\documentclass[]{fairmeta}

\usepackage{amssymb}

\usepackage{multirow}
\usepackage{tcolorbox}
\usepackage{verbatim}
\usepackage{multicol} 
\usepackage{circledsteps}
\usepackage{xspace}
\usepackage[T1]{fontenc}
\usepackage[utf8]{inputenc} 
\usepackage{textcomp}
\usepackage{tikz}
\usetikzlibrary{babel}
\newcommand{\sys}{SWE-Protégé\xspace}

\usepackage{tcolorbox}
\usepackage{capt-of}
\definecolor{chart}{HTML}{1f77b4}

\newtcolorbox{example}[1][]{
  colback=chart!5!white,
  colframe=chart,
  floatplacement=floating,
  title=\centering \textsf{#1}
}

\DeclareRobustCommand{\circled}[1]{\tikz[baseline=(char.base)]{
            \node[shape=circle,draw,inner sep=1pt] (char) {#1};}}

\title{\sys: Learning to Selectively Collaborate With an Expert Unlocks Small Language Models as Software Engineering Agents}

\author[\dagger,2]{Patrick Tser Jern Kon}
\author[1]{Archana Pradeep}
\author[2]{Ang Chen}
\author[1]{Alexander P. Ellis}
\author[1]{Warren Hunt}
\author[1]{Zijian Wang}
\author[3]{John Yang}
\author[1]{Samuel Thompson}

\affiliation[1]{Meta}
\affiliation[2]{University of Michigan}
\affiliation[3]{Stanford University}

\contribution[\dagger]{Work done at Meta}

\abstract{Small language models (SLMs) offer compelling advantages in cost, latency, and adaptability, but have so far lagged behind larger models on long-horizon software engineering tasks such as SWE-bench, where they suffer from pervasive action looping and low resolution rates. We introduce \sys, a post-training framework that reframes software repair as an expert--prot\'eg\'e collaboration problem. In \sys, an SLM remains the sole decision-maker while learning to selectively seek guidance from a strong expert model, recognize stalled states, and follow through on expert feedback. Our approach combines supervised fine-tuning on expert-augmented trajectories with agentic reinforcement learning that explicitly discourages degenerative looping and shallow expert usage. We lightly post-train Qwen2.5-Coder-7B-Instruct to achieve 42.4\% Pass@1 on SWE-bench Verified with SWE-agent, a +25.4\% improvement over the prior SLM state of the art, while using expert assistance sparsely ($\approx$4 calls per task and 11\% of total tokens).}

\date{\today}
\correspondence{\email{patkon@umich.edu}, \email{samuelt@meta.com}}

\begin{document}

\maketitle

\section{Introduction}
\label{sec:intro}

\begin{figure*}[t]
    \centering
    \includegraphics[width=\linewidth]{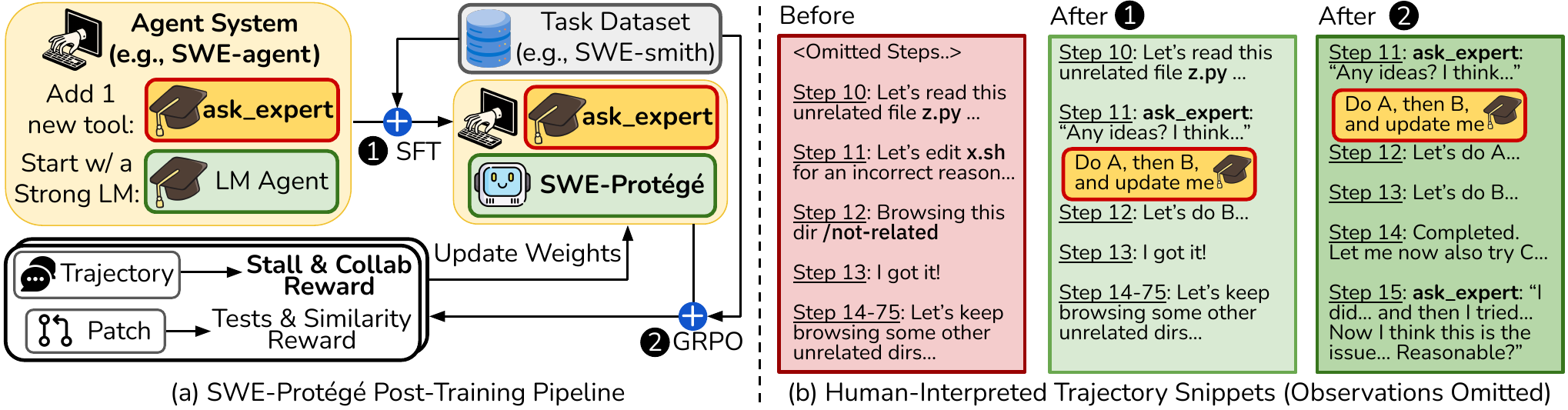}
    \caption{
(a) Our two-phase pipeline yields \sys-7B: Phase \circled{1} uses SFT on expert-augmented trajectories; Phase \circled{2} applies GRPO with trajectory-level rewards.
(b) Paraphrased trajectories. \textit{Before:} the SOTA SLM baseline (SWE-agent-LM-7B) fails to make reliable forward progress and degenerates into unproductive exploration. \textit{After Phase \circled{1}:} our SLM can invoke the expert, but follow-through on guidance is inconsistent and it often relapses into stalling. \textit{After Phase \circled{2}:} it learns to escalate when stalled, follow through on guidance, and report back, exhibiting multi-turn pair-programming behavior.
    }
    \label{fig:arch}
\end{figure*}

Software engineering (SE) tasks, most notably SWE-bench, have emerged as an important proving ground for language model (LM) agents, with recent advances driven primarily by large open-source and proprietary LMs (e.g., Claude Sonnet).
Yet this dominance stands in tension with a parallel trend~\citep{abouelenin2025phi,bakouch2025smollm3} toward small language models (SLMs; LMs with $\leq$10B parameters)~\citep{belcak2025small}, 
which offer compelling
advantages in cost, latency, and adaptability, but have so far remained largely absent from recent SWE-bench progress.

Indeed, this is unsurprising given the long-horizon/context nature of SE, which requires sustained 
progress over many turns (e.g., performing edits, recovering from errors) across complex repositories, rather than isolated single-turn code synthesis (e.g.,  HumanEval-style problems) where SLMs already perform reasonably well~\citep{hui2024qwen2}.
Even recent data-scaling approaches that generate large volumes of high-quality agent trajectories for training~\citep{wei2025swerladvancingllmreasoning, pan_training_2024}, including the state-of-the-art (SOTA) SWE-smith~\citep{yang2025swe}, remain insufficient for SLMs.
Prior work shows that SLM agents post-trained with these methods still suffer from stalled progress, which commonly manifests as degenerative action loops (i.e., SLM repeatedly calling the same base commands), and 
correspondingly 
poor resolution rates ($\approx$10\% Pass@1) on SWE-bench Verified~\citep{pan_training_2024, yang2025swe}.
Moreover, our analysis shows that progressively scaling SWE-smith training can even induce performance regressions.

Motivated by these limitations, we explore a complementary paradigm 
in \sys
that reframes the problem as one of expert–protégé
collaboration, 
analogous to pair programming,
leveraging the respective strengths of both components.
We study one effective instantiation of this paradigm in which 
the protégé (SLM) retains responsibility for the primary workflow and decision-making (handling most routine reasoning and interaction) with speed and low cost, while \textit{learning} to selectively seek guidance 
from
a strong expert (e.g., Claude Sonnet) that provides expensive but high-value intuition/understanding only when needed.

Specifically, \sys post-trains an SLM to act as an autonomous pair programmer that can 
(i) recognize stalled states where it's not making progress, (ii) invoke an expert appropriately, and (iii) execute and follow through on expert guidance over multiple turns.
We implement this in two phases (Fig.~\ref{fig:arch}). 
First, we distill into the SLM the mechanics and semantics of expert interaction via supervised fine-tuning on expert-augmented trajectories generated by a strong LM that is also repurposed to serve as the expert. 
Second, we run on-policy agentic reinforcement learning 
with shaped rewards that target when to escalate and how to collaborate over multiple turns:
the expert is repurposed as an in-trajectory judge (providing process-level supervision), which first explicitly discourages degenerative stalling, and then penalizes low-quality collaboration. 
Throughout, the SLM remains the primary decision-maker, treating the expert as an interactive collaborator rather than an oracle.

On SWE-Bench Verified, our \sys prototype lightly post-trains Qwen2.5-Coder-7B-Instruct to achieve 42.4\% accuracy, (+25.4\% over the prior state of the art, SWE-agent-LM-7B by SWE-smith). The model invokes expert assistance sparsely (only about 4 calls per task) 
with expert tokens comprising only $\sim$11\% of total tokens,
yielding up to 8.2$\times$ lower cost than an expert-only agent, signifying strong performance at low cost while preserving the SLM's autonomy. All results use the standard SWE-agent setup (75-step limit, no test-time sampling). Notably, these gains are achieved with minimal data and compute: we use the same SFT task dataset size ($\approx$5K trajectories) as SWE-smith, and perform a short single-node RL phase with rollouts generated from only 100 tasks for 160 steps. 
We envision a \sys-like paradigm shining in settings where, for instance, access to frontier-model tokens is a binding constraint: e.g., in application services~\citep{cursor, openclaw} constrained by cost, quotas, or rate-limits. 
\section{\sys Post-training Pipeline}
\label{sec:system}

\begin{table*}[t]
\centering
\renewcommand{\arraystretch}{1.15}
\setlength{\tabcolsep}{6pt}
\begin{tabular}{llcccc}
\toprule
\textbf{Model} & \textbf{System} & \textbf{Train Size} & \textbf{Accuracy (\%)} & \textbf{Reference} \\
\midrule
\multicolumn{5}{c}{\textit{Closed Weight Models}} \\[2pt]
\midrule
GPT-4o & OpenHands & - & 22.0 & \cite{openai2024gpt4ocard} \\
GPT-4o & SWE-agent & - & 23.0 & \cite{openai2024gpt4ocard} \\
Claude 3.5 Sonnet & Agentless & - & 50.8 & \cite{anthropicclaude35} \\
Claude 3.5 Sonnet & AutoCodeRover & - & 46.2 & \cite{anthropicclaude35} \\
Claude 3.5 Sonnet & OpenHands & - & 53.0 & \cite{anthropicclaude35} \\
Claude 3.7 Sonnet & SWE-agent & - & 58.2 & \cite{anthropicclaude37} \\
Claude 4.5 Sonnet & SWE-agent & - & \textbf{72} & - \\
Llama3-SWE-RL-70B & Agentless & 11M & 41.0 & \cite{wei2025swerladvancingllmreasoning} \\
\midrule
\multicolumn{5}{c}{\textit{Open Weight Models (Post-Trained)}} \\[2pt]
\midrule
Lingma-SWE-GPT-72B & SWE-SynInfer & - & 28.8 & \cite{ma2024lingma} \\
SWE-fixer-72B & SWE-Fixer & 110k & 32.8 & \cite{xie2025swe} \\
SWE-gym-32B & OpenHands & 491 & 20.6 & \cite{pan_training_2024} \\
R2E-Gym-32B & OpenHands & 3.3k & 34.4 & \cite{jain2025r2e-gym} \\
SWE-agent-LM-32B & SWE-agent & 5k & \textbf{40.2} & \cite{yang2025swe} \\
\midrule
\multicolumn{5}{c}{\textit{Open Weight Small Language Models ($\leq$10B parameters~\citep{belcak2025small})}} \\[2pt]
\midrule
SWE-gym-7B & OpenHands & 491 & 10.6 & \cite{pan_training_2024} \\
SWE-agent-LM-7B & SWE-agent & 2.4k & 17 & \cite{yang2025swe} \\
Lingma-SWE-GPT-7B & SWE-SynInfer & - & 18.2 & \cite{ma2024lingma} \\
\textbf{\sys-7B} (Sonnet 3.7) & SWE-agent & 4.9k & 30.8 &  \\
\textbf{\sys-7B} (Sonnet 4.5) & SWE-agent & 4.9k & 41.0 &  \\
\textbf{\sys-7B} (Opus 4.1) & SWE-agent & 4.9k & \textbf{42.4} &  \\
\bottomrule
\end{tabular}
\caption{Resolve rates (Pass@1) on SWE-bench Verified for existing solutions, taken directly from~\citep{yang2025swe}, compared to \sys.
We exclude open-weight models 
trained end-to-end from scratch~\citep{copet2025cwm}, 
and systems that rely on verifiers or test-time sampling.
The 17\% result corresponds to our strongest internally reproduced SWE-agent-LM-7B baseline.}
\label{tab:swe-result-compare-sota}
\end{table*}

Motivated by the aforementioned limitations, 
we introduce 
a post-training pipeline that enables an SLM to engage in \emph{autonomous, selective, and multi-turn collaboration} with an expert model during long-horizon tasks. 
\sys trains the SLM itself to (i) recognize stalled states, (ii) initiate escalation when appropriate, and (iii) follow through on expert guidance in a structured, iterative manner—treating the expert as an interactive collaborator rather than a one-shot oracle, analogous to pair programming initiated by the junior partner. 

\noindent\textbf{Problem Setting.}
We consider an agentic coding environment in which an SLM is embedded within an agent scaffold (e.g., SWE-agent~\citep{yang2024swe}) and prompted over multiple turns to solve a task (e.g., from SWE-bench).
At each turn, the SLM selects an action by invoking one of the agent’s tools---such as issuing shell commands (e.g., \texttt{grep}) or editing files---which are executed in a task-specific environment (e.g., a cloned GitHub repository).
Action selection is usually conditioned on a constructed representation of the interaction history (e.g., prior actions and tool outputs).
Upon termination, task success is assessed via external verification (e.g., unit tests in SWE-bench-style setups).

\subsection{Agent System Setup}
\label{subsec:system-setup}

\noindent\textbf{Agent Tool and Expert Interfaces.}
\sys requires only the addition of a single expert-invocation tool to an existing agent system.
Let \( s \in \mathcal{S} \) denote the full agent state, including the interaction history, tool outputs, and system prompts.
The SLM defines a policy \( \pi_\theta(a \mid s) \) over actions \( a \in \mathcal{A} \), where actions correspond to tool invocations.
We extend the action space to
\( \mathcal{A}' = \mathcal{A} \cup \{\texttt{ask\_expert}\} \)
where \texttt{ask\_expert} is a structured tool call in which the SLM formulates a query (e.g., a question). The expert processes this query and returns textual guidance, which is appended to the agent state \( s \) and subsequently used for continued decision-making.
In practice, all actions—including expert calls—are generated as text under a unified model distribution \( p_\theta(y \mid s) \), and expert invocation is detected by the agent controller via a reserved tool-call format.

The expert model is invoked with the agent’s query together with a system prompt (App.~\ref{app:expert-details}).
To keep escalation lightweight, the expert may be given only a partial view of \( s \);
specifically, it receives a compact summary \( \tilde{s} \) consisting of the most recent \( K \) interaction turns and relevant tool outputs.
In contrast, the agent always acts from the complete state \( s \).
We show later (\S\ref{sec:results}) that this asymmetric information flow keeps expert interactions focused and token-efficient.

\noindent\textbf{Collaboration Mechanisms.} 
Expert invocation occurs through two mechanisms. First, the SLM may autonomously emit \texttt{ask\_expert} as part of its policy. This learned invocation behavior is the primary focus of \sys; while training or adapting (e.g., prompt optimization~\citep{khattab2024dspy}) expert models is equally important and will likely yield even stronger performance, we assume that prompting strong frontier models yields sufficiently strong experts for our setting, and defer expert optimization to future work (\S\ref{sec:discussion}). Second, we support optional rule-based intervention (e.g., keyword triggers) that can forcibly invoke the expert. We currently use this for ablation purposes (\S\ref{subsec:result-ablation}).

\subsection{Phase I: Supervised Induction of Expert Usage}
\label{subsec:sft}

The first training phase equips the SLM with an operational understanding of expert interaction. Empirically, we find that SLMs fail to generalize to new tools (even simple ones) without explicit imitation learning. 
Crucially, this phase addresses \emph{how to interact} with the expert at the level of
mechanics and local semantics; e.g., invoking the tool, formulating contextually appropriate queries. It does not determine \emph{when} escalation is appropriate,
nor does it enforce correct downstream behavior after advice is given. 

\noindent\textbf{Synthetic Trajectory Generation.}
Training trajectories are generated using a strong code-capable model \(\pi_E\) conditioned on \(s\) with augmented action space \(\mathcal{A}'\),
in the same agent environment used for SLM inference. 
The model is prompted to invoke the expert tool only when appropriate (App.~\ref{app:synthetic-data}). 
In practice, this yields a natural data mixture in which expert invocation is sparse yet present in most trajectories, alongside a smaller subset with no expert calls, without any manual data mixing.
We also found it beneficial to instantiate the expert as the same model \( \pi_E \), and to additionally grant the expert access to the ground-truth solution (while explicitly instructing it not to reveal the answer verbatim).
This produces a substantially larger number of correct trajectories.\footnote{At any other point in time, including during evaluation, the expert no longer has access to the gold patch and unit tests and operates solely on \( \tilde{s} \). Details of prompting, filtering, and data statistics are provided in App.~\ref{app:synthetic-data}.}

\noindent\textbf{Supervised Fine-Tuning.}
We perform rejection-sampling supervised fine-tuning (SFT) on the expert-augmented trajectories. Given state–token pairs \( \{(s_i, y_i)\}_{i=1}^N \), we minimize the standard next-token cross-entropy loss:
\[
\mathcal{L}_{\mathrm{SFT}}(\theta)
= - \mathbb{E}_{(s_i, y_i)} \left[ \log p_\theta(y_i \mid s_i) \right].
\]
We introduce no auxiliary losses (e.g., 
regularizers that incentivize expert calls).
Instead, the SLM learns to sparsely invoke the expert in-context, emerging implicitly through imitation of trajectories where expert calls are infrequent.

\subsection{Phase II: RL for Expert--SLM Pair Programming}
\label{subsec:rl}

Phase~II 
(full details in App.~\ref{app:phase2-reward})
aligns the post-SFT SLM to behave as an autonomous pair programmer: it should
(i) \emph{escalate when progress stalls} instead of looping, and
(ii) \emph{follow through and report back} on expert guidance across multiple turns.
We achieve this with agentic RL and a composite reward that, in addition to standard objectives such as correctness, explicitly models (a) degenerative loops and
(b) 
low-quality collaboration.
Starting from the SFT checkpoint, we perform on-policy agentic RL with a GRPO-style objective (details in \S\ref{subsec:training-setup}). 

\noindent\textbf{Collaboration Quality.}
Let \(\tau\) denote the full agent trajectory (sequence of agent actions and tool responses).
During \(\tau\), the agent may call the expert \(N\) times; for call \(i\), let \(q_i\) be the agent query, \(g_i\) the expert guidance, and \(\tilde{s}_i\) the compact context packet shown to the expert (e.g., containing recent agent messages).
We associate each \(g_i\) with the subsequent \emph{response segment}
\(\Delta_i(\tau)\) 
(the agent’s actions and tool outputs after receiving \(g_i\) through the next expert call or termination).

We repurpose the expert itself as an \emph{in-trajectory judge} (akin to a value model~\citep{yuan2024self}) by defining two bounded functionals:
\begin{equation}
u_i := J_{\text{warrant}}(q_i,\tilde{s}_i) \in [0,1], \qquad
f_i := J_{\text{follow}}(g_i,\Delta_i(\tau)) \in [0,1].
\label{eq:judge-scores}
\end{equation}
where \(u_i\) scores whether escalation was warranted (discouraging lazy invocations),
and \(f_i\) scores whether the agent followed guidance and reported back (enforcing multi-turn
collaboration). These scores are computed via a hidden judge call and logged per event.

\noindent\textbf{Reward structure and gating.}
We use a trajectory-level composite reward inspired by~\citep{huang2025reinforcement}.
\begin{align}
R_{\text{total}}(\tau,x)
&=
R_{\text{loop}}(\tau)
+
w_{\text{follow}}\,R_{\text{follow}}(\tau)
\;+\;
g_{\text{loop}}(\tau)\,g_{\text{follow}}(\tau)\,R_{\text{other}}(\tau,x),
\label{eq:rtotal-main}
\\
R_{\text{other}}(\tau,x)
&=
R_{\text{correct}}(\tau,x)
+
w_{\text{sim}}\,R_{\text{sim}}(\tau,x)
\;+\;
w_{\text{expert}}\,R_{\text{expert}}(\tau).
\label{eq:rother}
\end{align}
Here \(x\) denotes the task instance metadata (e.g., gold patch), 
\(R_{\text{loop}}\le 0\) penalizes degenerative looping (command repetition); \(R_{\text{expert}}\)
shapes deferral quality using \(\{u_i\}\) (and anti-spam penalties); and \(R_{\text{follow}}\)
shapes follow-through using \(\{f_i\}\).
The gates \(g_{\text{loop}},g_{\text{follow}}\in\{0,0.5,1\}\) downweight \(R_{\text{other}}\)
under severe looping or failed follow-through, preventing correctness/similarity (derived from~\citep{copet2025cwm}) from masking
pathological interaction behavior.

\noindent\textbf{Reward shaping curriculum.}
To amplify the reward signal for the desired behaviors (i) and (ii), we adopt a two-stage reward shaping~\citep{ng1999policy} schedule:
(1) \emph{loop aggressive} shaping, which makes 
\(R_{\text{loop}}\) and \(g_{\text{loop}}\)
strongly suboptimal, anchoring stalled progress to expert invocation rather than continued unproductive exploration; and
(2) \emph{loop+follow aggressive} shaping, which additionally makes 
\(w_{\text{follow}}\) and \(g_{\text{follow}}\)
strongly suboptimal, encouraging the SLM to follow expert guidance to completion and report back, thereby inducing true multi-turn pair programming.
We specify the exact schedule (weights, caps, and gate thresholds) in
App.~\ref{app:phase2-shaping}.
As we show in \S\ref{sec:results}, these behaviors emerge after only a small number of RL updates on modest hardware, yielding further 
performance and efficiency
gains over \S\ref{subsec:sft}.
\section{Experimental Setup}
\label{sec:exp-setup}

We aim to explore the utility of our approach
(\S\ref{sec:system}) under a 
setting that is closely comparable to SWE-smith~\citep{yang2025swe}, the SOTA at the time of evaluation.
Across all experiments, we used a single node equipped with 8 NVIDIA A100/H100 80G GPUs
(e.g., AWS \texttt{p4de.24xlarge}) for both training and inference.

\noindent\textbf{Agent System.}
We adopt the off-the-shelf SWE-agent framework exactly as used in SWE-smith,
augmented with a single additional tool, \texttt{ask\_expert}, subject to a
75-step and \$2 budget. At each step, SWE-agent prompts the language model to produce a ReAct-style~\citep{yao2022react} \textit{(thought, action)} pair, where actions correspond either to file edits or shell commands within the task environment. Only 6 expert calls are permitted throughout all task rollouts.

\noindent\textbf{Models.}
Our base model is Qwen-2.5-Coder-Instruct~\citep{hui2024qwen2}, primarily
the 7B variant, as used in SWE-smith. We additionally report selected ablations with the 32B variant. Inference and rollout generation are performed using vLLM.
Expert models are accessed via AWS Bedrock and include Claude Sonnet 3.7, Sonnet 4.5, and Opus 4.1. 
Experts receive only the most recent 5 messages as context.

\noindent\textbf{Evaluation.}
We evaluate on SWE-bench Verified~\citep{chowdhury2024introducing}, a human-vetted
500-instance SWE-bench~\citep{jimenez2023swe} subset (drawn from 12 real-world GitHub repositories) to reduce ambiguity in problem statements and improve evaluation reliability. We report \%Resolved (Pass@1), i.e., the fraction of tasks solved by a single rollout per instance. We do not use multi-attempt sampling, majority voting, or other test-time scaling.
To assess potential data contamination of our approach, we additionally evaluate on a held-out subset of the SWE-smith task dataset 
comprising much newer tasks 
that is disjoint from all training data (details in \S\ref{subsec:result-ablation}).

\subsection{Training Setup}
\label{subsec:training-setup}

\noindent\textbf{SFT Details.}
We perform rejection-sampling full SFT on the Qwen-2.5-Coder-Instruct-7B base model with Torchtune~\citep{torchtune} using the same SWE-smith task dataset~\citep{swe_smith_dataset}, for a fair comparison.
Specifically, we train on trajectories generated by Claude Sonnet 3.7 using our procedure (\S\ref{subsec:sft}; which yielded 38\% more usable trajectories than when the ground-truth patch was not provided) by sampling until we obtained $\approx$4.8K resolved tasks.
We use a maximum sequence length of 32{,}762 and batch size 32.
We also experimented with LoRA and QLoRA under the same data and evaluation protocol; both underperformed full SFT on SWE-bench Verified Pass@1. 
Thus, we report full SFT results throughout.

\noindent\textbf{RL Details.}
Starting from the SFT checkpoint, we apply an on-policy RL stage using GRPO~\citep{khatri2025art}. While stronger variants exist~\citep{yu2025dapo}, we only use a minimal modification to standard GRPO: asymmetric clipping (following~\citep{yu2025dapo}) with separate upper/lower clip thresholds, setting $\epsilon_{\text{high}}=0.28$ and $\epsilon_{\text{low}}=0.20$  to reduce premature entropy collapse.
Concretely, for a prompt $x$, we sample $G$ completions $\{a_i\}_{i=1}^G \sim \pi_{\theta_{\text{old}}}(\cdot \mid x)$ and optimize
\[
J(\theta)=\frac{1}{G}\sum_{i=1}^{G}\min\!\Big(r_i(\theta)A_i,\,\mathrm{clip}\!\big(r_i(\theta),\,1-\epsilon_{\text{low}},\,1+\epsilon_{\text{high}}\big)A_i\Big)-\beta\,\mathrm{KL}\!\big(\pi_\theta(\cdot\mid x)\,\|\,\pi_{\mathrm{ref}}(\cdot\mid x)\big).
\]
where $r_i(\theta)=\pi_\theta(a_i\mid x)/\pi_{\theta_{\text{old}}}(a_i\mid x)$.
The advantage is computed via group normalization,
\[
A_i=\frac{r_i^{\text{env}}-\operatorname{mean}(\{r_j^{\text{env}}\}_{j=1}^G)}
{\operatorname{std}(\{r_j^{\text{env}}\}_{j=1}^G)},
\]
with $r_i^{\text{env}}$ denoting the scalar rollout reward.
We train on a 100-task subset drawn from SWE-Gym~\citep{SWE_Gym_Subset} (bundled with SkyRL), with no dataset mixing. We used 6 rollouts per prompt, batch size 16, and 160 total steps. This RL phase is intentionally short and data-light, targeting behavioral shaping of expert usage and trajectory hygiene rather than broad capability gains.

Our reward shaping schedule is (details in App.~\ref{app:phase2-reward}):
(i) \emph{loop aggressive} shaping for steps 1--80, where we set
\((k_1,k_2,\lambda_{\text{loop}},c_{\text{loop}})=(15,8,0.5,-10)\)
and only activate the \(g_{\text{loop}}\) gate.
(ii) \emph{loop+follow aggressive} shaping for steps 81--160, where we retain the same loop penalty, set
\((w_{\text{expert}},w_{\text{follow}})=(0.3,2.0)\) and
\((\tau_{\text{follow}},p_{\text{follow-low}})=(0.5,-2.0)\),
activate the \(g_{\text{follow}}\) gate, and impose a hard \(-10\) penalty when no expert call is made.

Our RL system builds on Ray-based SkyRL~\citep{skyrl}. We integrate SWE-agent by implementing a SkyRL \emph{generator} that, for each sampled task, launches a Ray worker to run a full SWE-agent episode in a SWE-ReX Docker runtime. To improve stability and throughput, we (i) cap concurrent SWE-agent/Docker startups and introduce I/O backoff to mitigate stalls and transient failures, (ii) add trajectory-level checkpointing of completed rollouts to support mid-run reward-shaping updates, and (iii) pipeline inference with multiple in-flight batches to maintain high vLLM utilization despite stragglers.
\section{Results}
\label{sec:results}

\begin{figure*}[t]
    \centering
    \begin{minipage}[t]{0.49\textwidth}
      \centering
      \includegraphics[width=\linewidth]{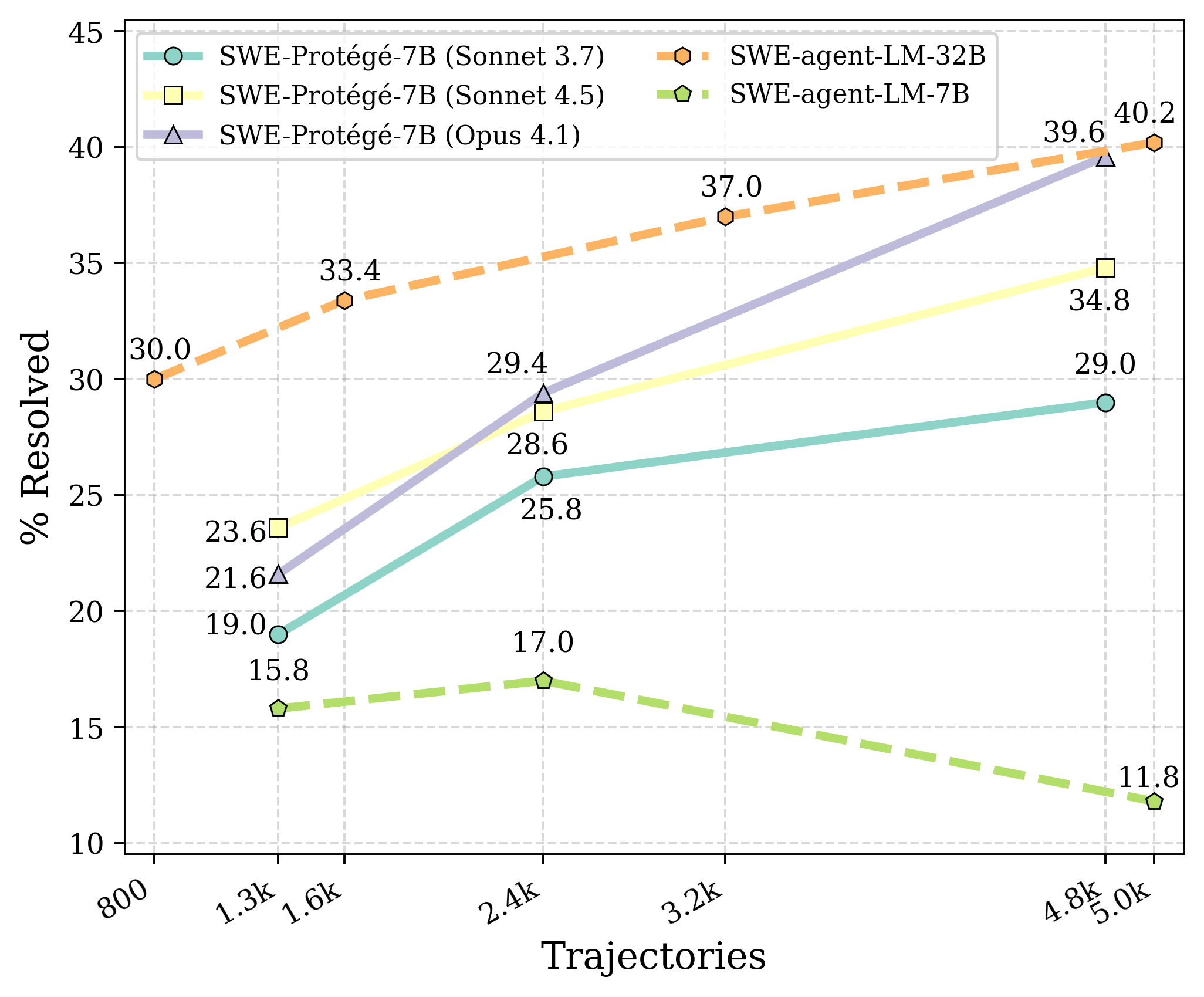}
      \captionof{figure}{\sys-7B exhibits consistent gains via SFT with increased SWE-smith training data.}
      \label{fig:main-results}
    \end{minipage}
    \hfill
    \begin{minipage}[t]{0.49\textwidth}
      \centering
      \includegraphics[width=\linewidth]{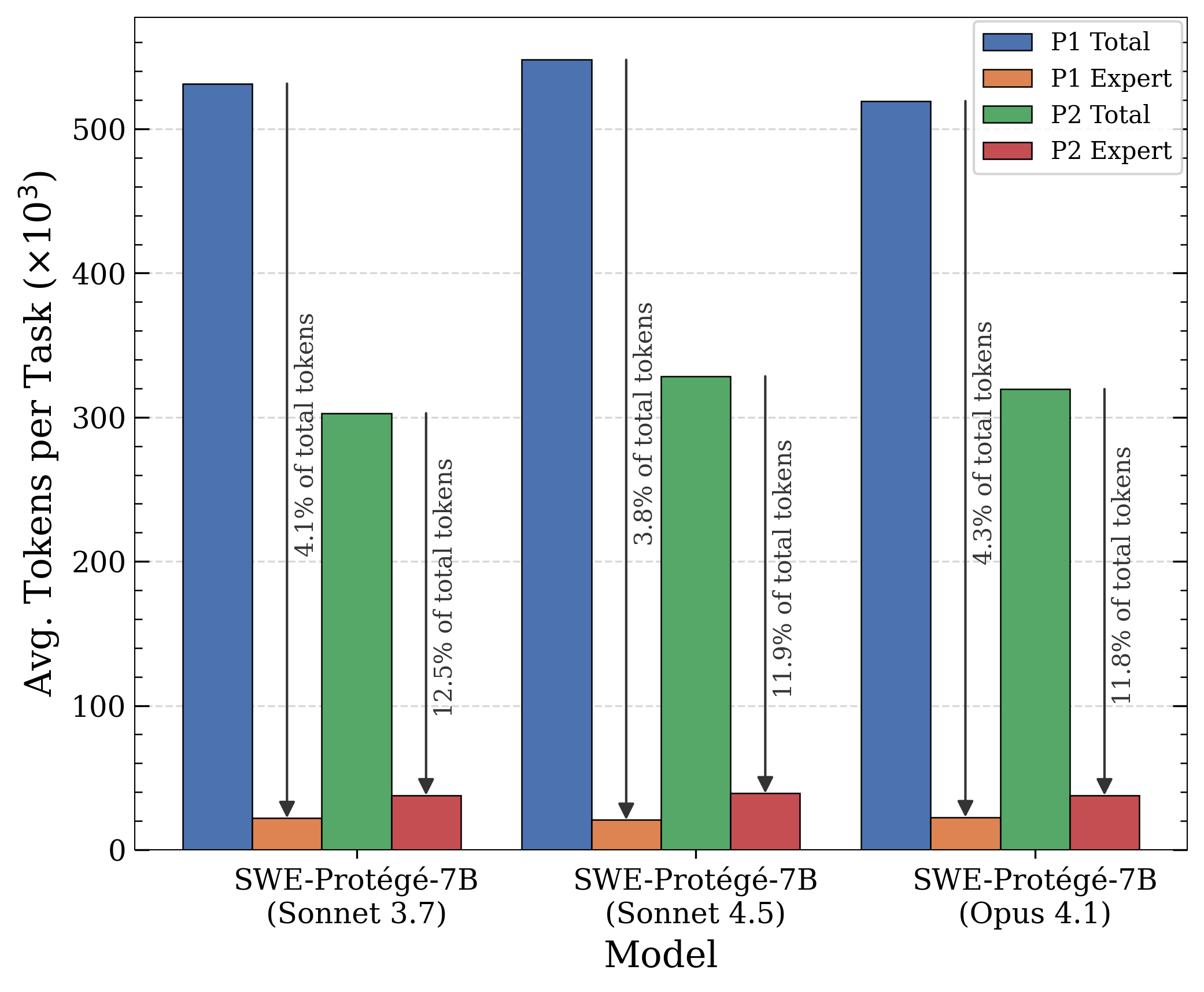}
      \captionof{figure}{Expert tokens remain consistently low, while total token usage is substantially reduced after Phase II.}
      \label{fig:token_usage}
    \end{minipage}
\end{figure*}

\begin{figure*}[t]
    \centering
    \begin{minipage}[b]{0.52\textwidth}
      \vspace{0pt}
      \centering
      \includegraphics[width=\linewidth]{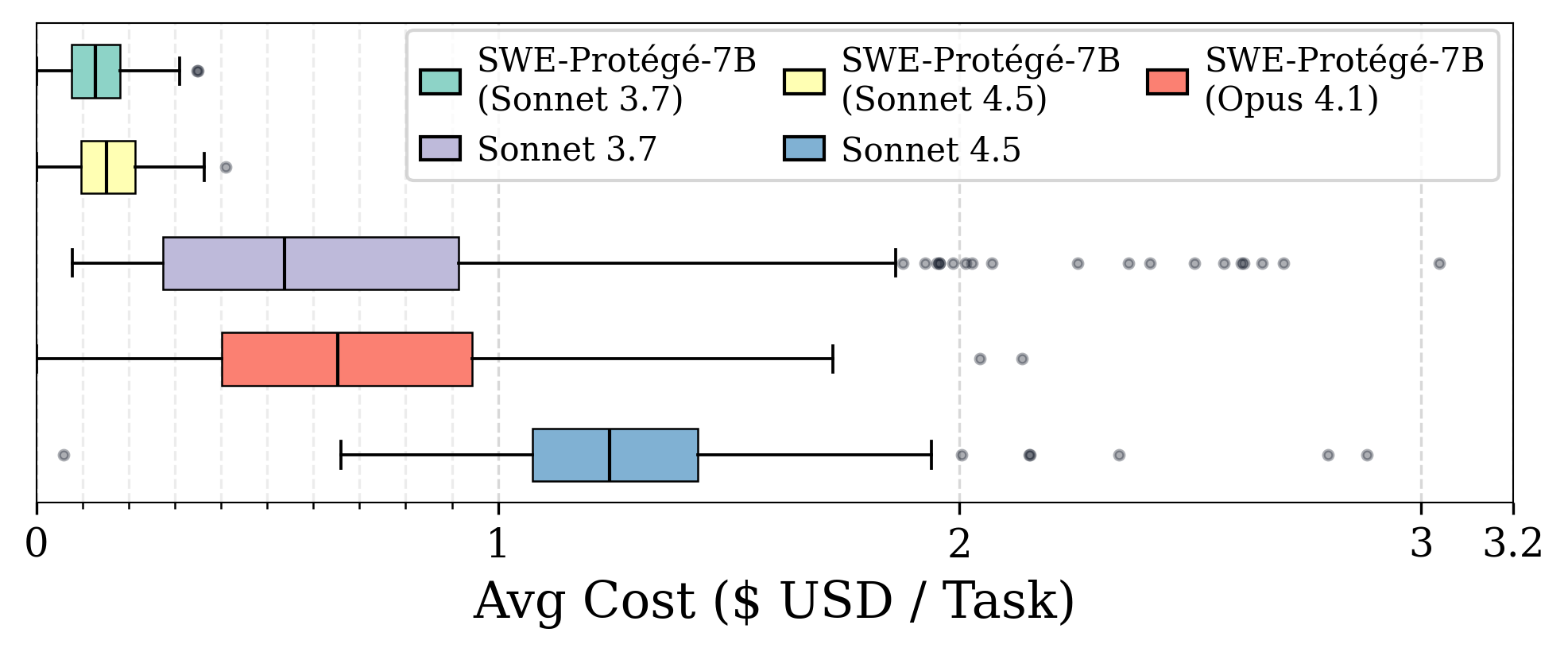}
      \captionof{figure}{\sys substantially reduces per-task cost relative to direct expert execution (the expert solves the entire task). Details in \S\ref{subsec:result-ablation}.}
      \label{fig:cost_distribution}
    \end{minipage}
    \hfill
    \begin{minipage}[b]{0.47\textwidth}
      \vspace{0pt}
      \centering
      \small
      \setlength{\tabcolsep}{6pt}
      \renewcommand{\arraystretch}{1.15}
      \begin{tabular}{lcc}
      \toprule
      \textbf{Model} & \textbf{Pass@1 (\%)} & $\Delta$ \\
      \midrule
      \sys-7B (Sonnet~3.7) & 30.6 & +1.2 \\
      \sys-7B (Sonnet~4.5) & 41.0 & \textbf{+6.2} \\
      \sys-7B (Opus~4.1)   & \textbf{42.4} & +2.8 \\
      \bottomrule
      \end{tabular}
      \vspace{35pt}
      \captionof{table}{Phase~II RL results in performance improvements on SWE-bench Verified. $\Delta$ denotes absolute improvement over the corresponding SFT checkpoint.}
      \label{tab:rl-improvement}
    \end{minipage}
\end{figure*}

Table.~\ref{tab:swe-result-compare-sota} summarizes our main results on SWE-bench Verified.
To ensure a fair comparison to SWE-smith (the SOTA open-weight scaling baseline) we restrict expert backends to the same evaluation lineage, using Sonnet~3.7 as in SWE-smith and additionally considering newer successors (Sonnet~4.5 and Opus~4.1), even though stronger options exist today.
\sys-7B achieves up to 42.4\% accuracy, exceeding the best prior open-weight scaling baseline SWE-agent-LM-32B (40.2\%) by +2.2\%, while substantially improving over the corresponding SLM baseline SWE-agent-LM-7B (+30.6\%) and the strongest prior 7B-class SWE system Lingma-SWE-GPT-7B (+24.2\%).
Finally, Phase~II RL training starting from the SFT checkpoints (Table~\ref{tab:rl-improvement}) consistently improves performance across expert backends, delivering an average gain of 3.4\%.

\noindent\textbf{Performance Scales with More Training Data.}
Fig.~\ref{fig:main-results} plots accuracy as a function of the number of training trajectories.
For each point, we subsample trajectories uniformly at random from our $\approx$4.8k source task pool and train from the same base initialization.
We also independently re-trained/evaluated the SWE-agent-LM-7B scaling curve using the \emph{original} trajectory pool~\citep{swe_smith_trajectories} and matching sampling protocol for each budget, and observed accuracy consistent with SWE-smith reports (e.g., $\approx$15--17\% at the 2k--2.4k regime).
We focus on the SWE-smith trajectory source because it was shown to provide stronger improvements than alternative sources (e.g., SWE-Gym~\citep{pan_training_2024}) at comparable dataset sizes.
For \sys-7B, increasing the SFT trajectory budget yields monotonic gains across all three experts: e.g., 
with Sonnet~3.7, accuracy improves from 19.0\% (1.3k) 
$\rightarrow$ 33.4\% (4.8k), a net gain of +14.4\%;
and with Sonnet~4.5, 23.6\% 
$\rightarrow$ 39.4\% (+15.8\%).
In contrast, SWE-agent-LM-7B exhibits a plateau from 2.2\% at 100 trajectories to 17.0\% at 2.4k, followed by regression at larger budgets (11.8\% at 5.0k); only the higher-budget regime is shown in Fig.~\ref{fig:main-results}.
These trends align with SWE-smith’s finding that high-quality trajectories can drive strong performance gains, but extend it by showing that such gains can be realized in SLMs when trajectories are augmented with sparse expert interactions; we note that direct prompting is insufficient to elicit this behavior (we did targeted ablations on SWE-smith’s trained 7B and 32B models).

\subsection{Ablations on \sys}
\label{subsec:result-ablation}

\noindent\textbf{Expert Collaboration is Token-light.}
Fig.~\ref{fig:token_usage} decomposes per-task generation into \sys tokens vs.\ expert tokens. 
We focus our discussion on Phase~II (P2), as the qualitative patterns are consistent across phases.
Expert tokens account for only a small fraction of each trajectory (e.g., 11.9\% with Sonnet~4.5), implying that \sys performs the bulk of reasoning, exploration, and patching, while using the expert as a sparse collaborator rather than a primary driver.
Expert replies are short (median / p95 / max: 500 / 937 / 1,657 tokens) despite being conditioned on large input contexts (8,885 / 20,716 / 43,031 tokens) consisting of the SLM query and five prior messages (\S\ref{sec:exp-setup}).
Meanwhile, total tokens per task remain $\approx$constant across experts 
($\sim (3\text{--}3.2)\times 10^5$), 
suggesting that swapping experts primarily changes \emph{quality} rather than the amount of agent-side work. 
Phase~II reduces total token usage by $\approx$40\%, primarily by eliminating degenerative looping.
\sys also yields large reductions in expert-related cost. From Fig.~\ref{fig:cost_distribution}, direct expert execution has median per-task costs of \$0.54 (Sonnet~3.7) and \$1.24 (Sonnet~4.5), with outliers reaching \$3.04 and \$2.88. Under \sys, the median expert cost drops to \$0.13/\$0.15 for Sonnet~3.7/4.5 (4.2$\times$ and 8.2$\times$ lower than direct execution), and to \$0.65 with Opus~4.1; despite Opus~4.1 being 5$\times$ and 4.54$\times$ more expensive per-token, it remains cost-comparable to direct Sonnet~3.7 and cheaper than direct Sonnet~4.5.
Finally, we separately test Sonnet~3.7 and Sonnet~4.5 on SWE-agent with step limits of 8 and 16, respectively, which is a more generous budget that exceeds both our 6-expert-call cap and average expert cost (by $\sim$1.47$\times$ and 1.03$\times$, respectively). We find performance drops sharply to 18.2\% and 26\%.

\begin{figure*}
   \centering
   \includegraphics[width=0.7\linewidth]{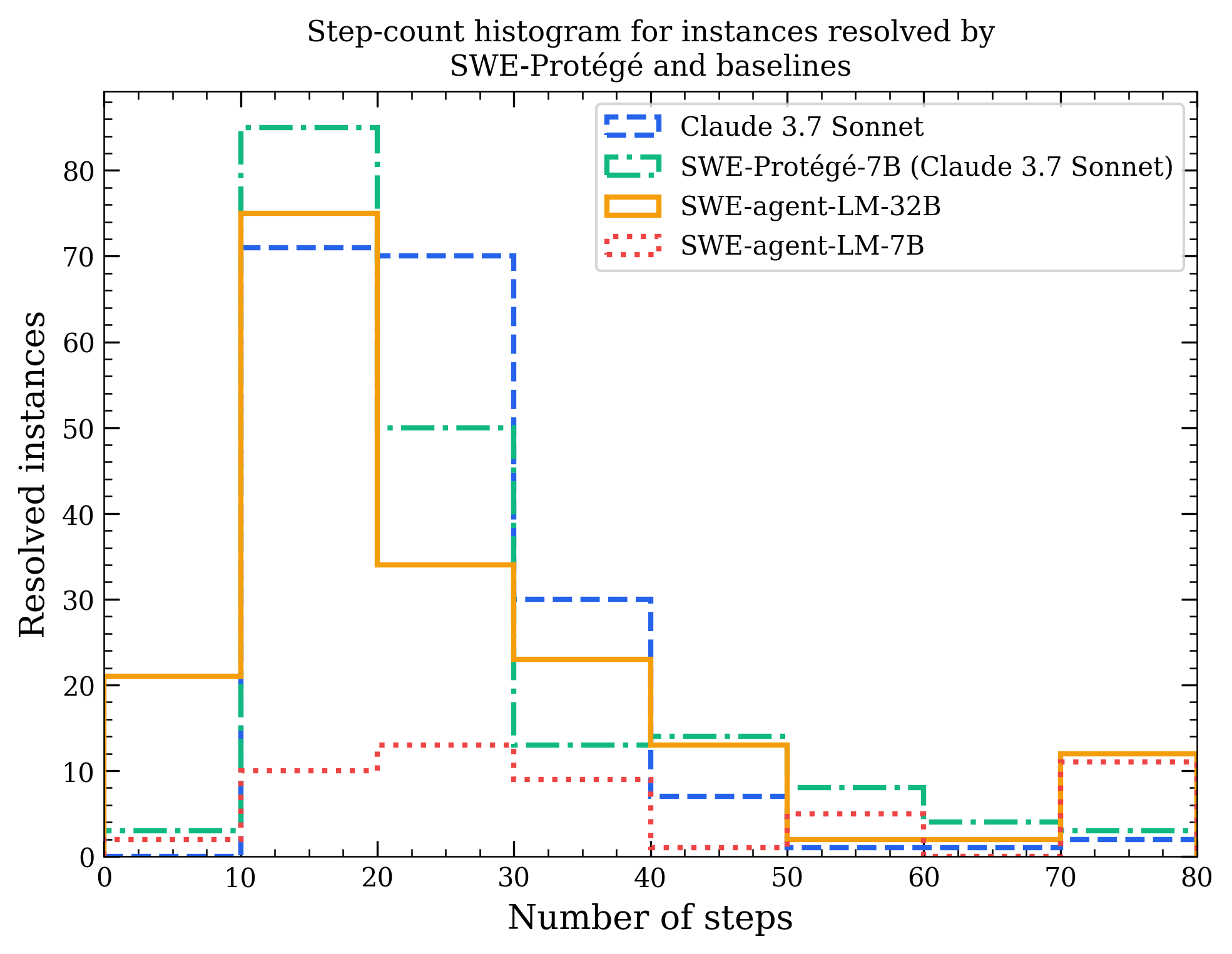}
   \caption{Agent behavior diagnostics adapted from~\cite{yang2025swe}. Histogram of the number of steps taken to complete the task per resolved trajectory.}
   \label{fig:step-count-histogram}
\end{figure*}

\noindent\textbf{\sys-7B Solves Tasks Efficiently.}
Beyond reducing expert token usage, \sys also becomes progressively more efficient at the \emph{trajectory level}. As training proceeds, the mean number of steps per task drops from $\approx$60 at the post-SFT checkpoint to $\approx$20 after Phase~II, while the number of expert calls remains stable at around four (Fig.~\ref{fig:mean-rl-reward}). Fig.~\ref{fig:step-count-histogram} further shows that \sys achieves step counts comparable to Sonnet~3.7 and SWE-agent-LM-32B.
This decoupling indicates that RL does not simply suppress collaboration, but instead trains \sys to collaborate \emph{more productively}, converging in fewer overall steps.
Importantly, shorter trajectories do not imply premature termination; \sys maintains the ability to remain focused in long-horizon settings: on average, 10.8\% of instances are resolved after $\geq$40 steps. 
Combined with our findings earlier, these effects explain the substantial cost savings observed in Fig.~\ref{fig:cost_distribution}.
\sys saves cost along two orthogonal dimensions: (i) minimizing expensive expert tokens, and (ii) shortening trajectories overall.

\begin{figure*}
    \centering
    \includegraphics[width=1\linewidth]{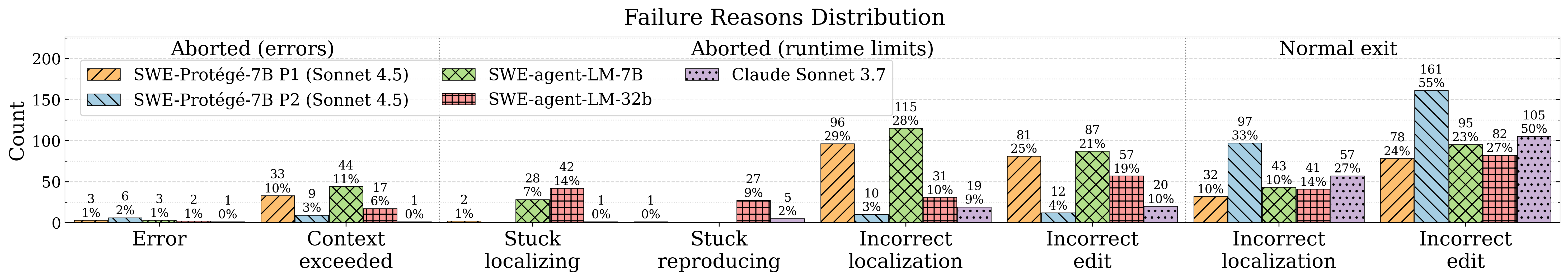}
    \caption{Phase~II RL (P2) sharply reduces cost/step limit aborts relative to post-SFT (P1) and SWE-smith baselines.}
    \label{fig:failure-reasons-eval}
\end{figure*}

\begin{figure*}[t]
    \centering
    \begin{minipage}[t]{0.44\textwidth}
      \vspace{0pt}
      \centering
      \includegraphics[width=\linewidth]{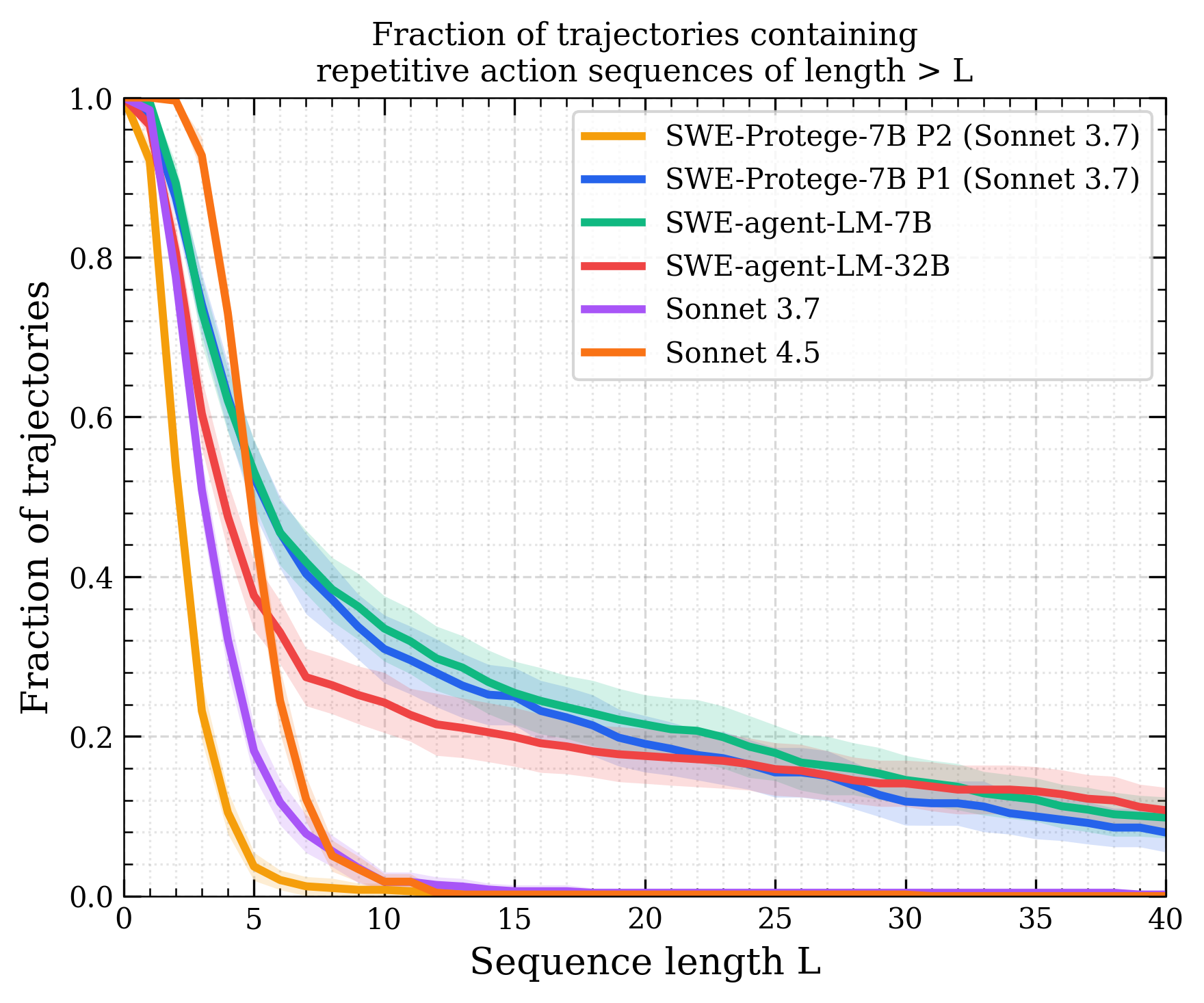}
      \vspace{-15pt}
      \captionof{figure}{Repetitive unproductive actions are largely eliminated in \sys-7B after Phase 2 RL, though they persist even after Phase 1 SFT.}
      \label{fig:loops}
    \end{minipage}
    \hfill
    \begin{minipage}[t]{0.55\textwidth}
      \vspace{10pt}
      \centering
      \includegraphics[width=\linewidth]{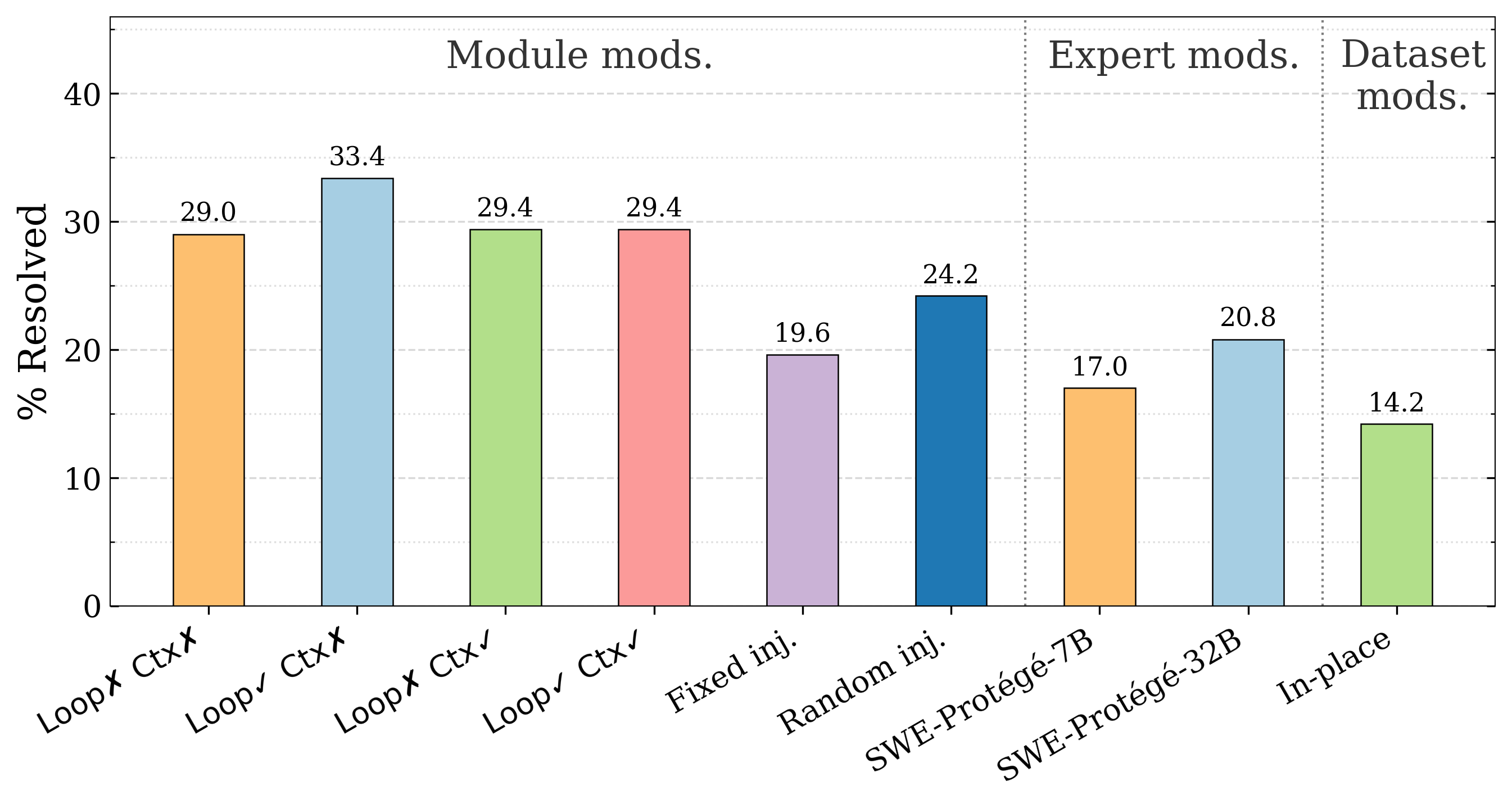}
      \captionof{figure}{\sys variants on SWE-bench Verified. Module modifications explore some alternative collaboration strategies. In ``Expert mods.'', we evaluate two lightly post-trained in-house experts. In ``Dataset mods.'', we approximate our SFT data generation process using in-place modifications to existing trajectories.}
      \label{fig:minicoder_variants}
    \end{minipage}
\end{figure*}

\noindent\textbf{Stalling is Replaced with Effective Collaboration.} 
Prior work has identified repetitive actions as a major failure mode for SLMs~\citep{pan_training_2024, yang2025swe}, e.g., models may repeatedly issue localization/inspection commands (e.g., search, open, grep) 
We quantify this behavior using repeated tool-use sequences: we define a \emph{loop} as a contiguous repetition of a tool-action pattern with length at least $L$, and report the fraction of trajectories that contain any loop longer than $L$.
Fig.~\ref{fig:loops} shows that \sys-7B post-SFT (P1) suffers from severe looping: 31.0\% of trajectories contain a repeated-action run longer than 10 steps (19.0\% $>20$; 8.0\% $>40$), comparable to SWE-agent-LM-7B/32B (33.6\%/24.4\% $>10$).
Phase~II RL (P2) sharply reduces this failure mode: only 0.8\% of trajectories have runs longer than 10 and none exceed 20, outperforming Sonnet~3.7 and 4.5 (both 1.8\% $>10$).
As seen in Fig.~\ref{fig:mean-rl-reward} and Fig.~\ref{fig:negative-rl-reward}, under our \emph{loop aggressive} shaping stage, loop violations collapse (mean loop penalty rises from $\approx-1.3$ toward $\approx0$; loop-negative rate drops from $\approx 0.50$--$0.65$ to near-zero by $\approx100$ global steps), yet SWE-bench accuracy improves only marginally.
Performance gains emerge only after switching to \emph{follow-aggressive} shaping, where the policy learns to request and execute expert guidance faithfully: the mean follow-through score increases from roughly $\approx0.2$ early in training to $\approx0.8$ by the end, and the fraction of trajectories with negative follow-through declines from $\approx0.30$--$0.40$ to $\approx0.05$.
This allows us to convert stalled states into forward progress.
We corroborate this via failure-mode shifts (Fig.~\ref{fig:failure-reasons-eval}).
Relative to post-SFT (P1), post-RL (P2) markedly reduces aborts due to runtime limits, a common failure even for SWE-agent-LM-32B, indicating that \sys no longer gets stuck. Instead, it decisively follows expert guidance through to completed end-to-end attempts, even when that guidance is imperfect.

\noindent\textbf{Ablations on \sys Variants.} 
We test a number of variants (Fig.~\ref{fig:minicoder_variants}) using \sys-7B P1. 
In \textit{Module mods.}, we study alternative expert--protégé collaboration strategies.
Specifically, we evaluate (i) a looping intervention heuristic that forcibly invokes the expert once a loop-length threshold (initial: 15, subsequent: 8) is exceeded (\textit{Loop}), and (ii) limiting the interaction history passed to the expert to only the 5 most recent messages  (\textit{Ctx}).
We find that enabling the loop heuristic improves performance only when the expert observes the full interaction history: Loop$\checkmark$ Ctx$\times$ achieves 33.4\%, compared to 29.0\% for Loop$\times$ Ctx$\times$ (+4.4\%).
However, with Ctx$\checkmark$, the loop heuristic provides no measurable benefit (29.4\% with or without Loop under Ctx$\checkmark$).
Moreover, variants that \emph{remove} the policy’s ability to autonomously request help and instead rely solely on passive expert invocation underperform sharply:
fixed-interval invocation drops to 19.6\%, and random invocation to 24.2\%, despite receiving comparable or even more frequent expert advice.
We view other collaboration strategies as an important direction for future work.

In \textit{Expert mods.},
we evaluate in-house experts initialized from our own \sys-7B and \sys-32B checkpoints and fully SFT’ed to serve as experts using a separate instruction-tuning dataset, which is derived from the expert-augmented trajectories (\S\ref{subsec:training-setup}): we treat the pre-expert interaction context as input and the expert message as the target, yielding 5{,}623 examples.
Both in-house experts underperform frontier backends, though increasing expert capacity still helps: replacing \sys-7B expert with \sys-32B expert improves task resolution from 17.0\% to 20.8\% under the same protégé policy.
We therefore treat expert post-training as an important but non-trivial and orthogonal direction, and leave a systematic exploration to future work.
Finally, in \textit{Dataset mods.}, we test whether \S\ref{subsec:sft}'s synthetic data generation (fresh expert-augmented rollouts) can be approximated by \emph{in-place} modifications of existing SWE-smith trajectories.
The in-place control—injecting additional expert calls into existing traces and minimally editing subsequent messages to acknowledge them—performs worst (14.2\%), even below the SWE-agent-LM-7B baseline (17.0\%).
This gap suggests that the gains from our data pipeline stem not from simply adding expert tokens, but from inducing coherent interactions that the protégé meaningfully conditions on.

\noindent\textbf{Contamination Study.}
Because \sys relies on external expert models, gains on SWE-bench Verified could in principle be inflated by data leakage or memorization.
To assess robustness, we evaluate on a held-out SWE-smith-style subset of 400 tasks that was explicitly excluded from the trajectory-generation mixture used to train our 7B model.
Importantly, this subset was released \emph{after} the expert models became available (e.g., Sonnet~3.7 was released on February~24,~2025, while SWE-smith was released on April~29,~2025), reducing the likelihood of overlap.
We follow the same interaction protocol as in the main evaluation: a single rollout with a fixed step/cost budget while furnishing the expert with the same compact context \(\tilde{s}_i\) (\S\ref{subsec:rl}); as before, the expert observes only truncated recent context and has no access to the gold patch or unit-tests.
Under this shifted evaluation, the 7B model achieves 32.0\% accuracy without expert calls and 40.3\% with Sonnet~3.7, while the 32B variant achieves 41.5\% without expert calls and 43.0\% with Sonnet~3.7, indicating that gains from
\sys
persist beyond the training distribution.

\section{Related Work}
\label{sec:related-work}

\noindent\textbf{Training Software Agents.}
SWE-Smith~\citep{yang2025swe} addresses data scarcity by synthesizing large numbers of verifiable software engineering task instances from 128 GitHub repositories.
Lingma-SWE-GPT~\citep{ma2024lingma} adopts a development-process-centric training scheme for 7B/72B variants; SWE-Gym~\citep{pan_training_2024} introduces an open training environment that improves agents at 7B/32B scales; SWE-Fixer~\citep{xie2025swe} trains specialized retriever and editor models for efficient issue resolution.
The above approaches rely on distilled data from frontier models (e.g., GPT or Claude), and post-train Qwen2.5 models via supervised finetuning.
SWE-RL~\citep{wei2025swerladvancingllmreasoning} applies RL to Llama~3~\citep{grattafiori2024llama}, requiring substantial data/compute (e.g., 273k seed tasks, 512 H100 GPUs) and a custom agent scaffold.
CWM~\citep{copet2025cwm} (32B) performs extensive end-to-end training, combining large-scale data/compute, custom agent scaffold, and test-time scaling to achieve strong SWE-bench performance.
In contrast, \sys focuses on lightweight post-training and demonstrates that an SLM can reach competitive performance relative to SOTA open-weight models. 

\noindent\textbf{Model Routing.}
Here, a router selects the most suitable LM to handle a query. Existing work has so far focused on \emph{per-task} routing and \emph{single-turn} tasks: e.g., \emph{non-predictive} routing, which executes models sequentially and escalates based on output evaluation until a quality threshold is met~\citep{chen2023frugalgpt}; and \emph{predictive} routing, which selects a model \emph{a priori} for a task using LM-based heuristics~\citep{ong2024routellm}, learned neural routers~\citep{jiang2023llm}, or cluster-based methods~\citep{jitkrittum2025universal,zhang2025avengers}.
In contrast, we study long-horizon, multi-turn agentic coding tasks,
where
defining a reliable per-step routing signal is 
ill-posed (e.g., a 
syntax error 
does not necessarily justify escalation). 
Moreover, we allow the participating LMs to \textit{self-determine when and how to collaborate}; in our instantiation, the SLM performs the bulk of routine reasoning and actions and experts are invoked selectively as needed.

\noindent\textbf{Small Language Models.}
SLMs are increasingly viewed as practical agents when the domain scope is narrow, offering advantages in inference efficiency (e.g., latency, memory footprint) and fine-tuning agility~\citep{belcak2025small}. Many use-cases~\citep{wang2025comprehensive} have emerged: e.g., single-turn question answering~\citep{abouelenin2025phi}, mathematical reasoning~\citep{guan2025rstar}, and single-turn coding tasks~\citep{bakouch2025smollm3}. In contrast, we developed the first usable SLM on a long-horizon, agentic coding task.

\section{Discussion}
\label{sec:discussion}

\noindent\textbf{Limitations and Future Work.}
While we focus on SWE-bench within the SWE-agent framework, \sys{}’s techniques are not tied to this setting and could in principle be applied to other domains (e.g., data analysis).
We do not exhaustively explore design choices: e.g., Phase~I/II hyperparameters, alternate 
collaboration strategies 
(e.g., expert interrupts or richer bidirectional control), or broader student model families; since our goal is to establish that \sys can materially improve SLM performance rather than fully optimize the frontier.
Finally, we treat the expert as a fixed black-box backend and leave more principled expert post-training and co-adaptation as an important future direction.

\noindent\textbf{Conclusion.}
We show that SLMs, while previously lagging on long-horizon software repair, can achieve strong SWE-bench performance when trained to collaborate effectively with an expert.
\sys lightly post-trains Qwen2.5-Coder-7B-Instruct to reach \textbf{42.4\%} Pass@1 on SWE-bench Verified (\textbf{+25.4\%} over the prior SLM SOTA), while using expert assistance sparsely ($\approx$11\% of total tokens).
Our results suggest that learned expert--prot\'eg\'e collaboration is a practical path for advancing effective, fast and cost-efficient SLM agents.
\section*{Impact Statement}

\sys aims to make long-horizon agentic systems more practical by enabling small language models to remain the primary decision-makers while selectively collaborating with stronger expert models when progress stalls. This design targets a more economical deployment model for agentic AI, where large models are used sparingly rather than continuously. In software engineering workflows, this can reduce latency and compute costs, and may make agentic assistance more accessible in settings where infrastructure or budget constraints preclude always-on frontier models.

\sys is not intended to replace human developers or engineering judgment. Instead, it supports a layered assistance paradigm in which a small model handles routine exploration and tool use, while expert models are invoked selectively to overcome uncertainty or stagnation. This mirrors common practice in production engineering, where automated systems rely on fallbacks, escalation paths, and checks before changes are merged or deployed.

Like other LLM-based code agents, \sys inherits risks associated with incorrect, misleading, or incomplete outputs. In our setting, these risks include improper deferral decisions (e.g., escalating too late, too frequently, or inappropriately) and propagating incorrect guidance from the expert model. Misguided edits could introduce defects, regressions, or security vulnerabilities if deployed without adequate review. We therefore recommend deploying \sys with automated verification (e.g., unit tests, linters, static analysis) and human oversight, particularly in safety-critical or production environments.

As with other LLM-based code agents, \sys is subject to biases arising from benchmark composition, training data, and model interaction dynamics. Our evaluation focuses primarily on Python-based repositories, reflecting SWE-bench's task composition. To mitigate expert interaction bias, expert models are treated as fallible collaborators rather than oracles: they have no access to ground-truth patches or tests during evaluation/inference, and the SLM remains the principal decision-maker. We further mitigate hidden biases through the use of public datasets (e.g., SWE-smith), standard evaluation protocols (i.e., from SWE-bench), and transparent reporting of training procedures, reward design, and limitations, which we explicitly document to avoid overgeneralization of results.

Our experiments and analyses are conducted on open-source repositories and public benchmarks. Lowering the cost of code generation and repair may increase the volume of automated changes, which can amplify downstream risks if used irresponsibly. Careful access control, auditability, and review processes remain important. 
Except for the SWE-agent-LM 7B and 32B variants and the Claude Sonnet 3.7, Sonnet 4.5, and Opus 4.1 models, which we evaluate as baselines or experts in this work, all other entries in Table~\ref{tab:swe-result-compare-sota} are taken directly from~\citep{yang2025swe} and were not evaluated by us.

Overall, we view \sys as a step toward more responsible and sustainable agentic systems: ones that use scale selectively, incorporate explicit escalation mechanisms, and emphasize controlled interaction with stronger models rather than ubiquitous reliance on them.

\clearpage

\bibliographystyle{assets/plainnat}
\bibliography{paper}

\clearpage
\newpage
\beginappendix
\section{Extended Evaluation}

\begin{figure}[t]
\centering
\includegraphics[width=\linewidth]{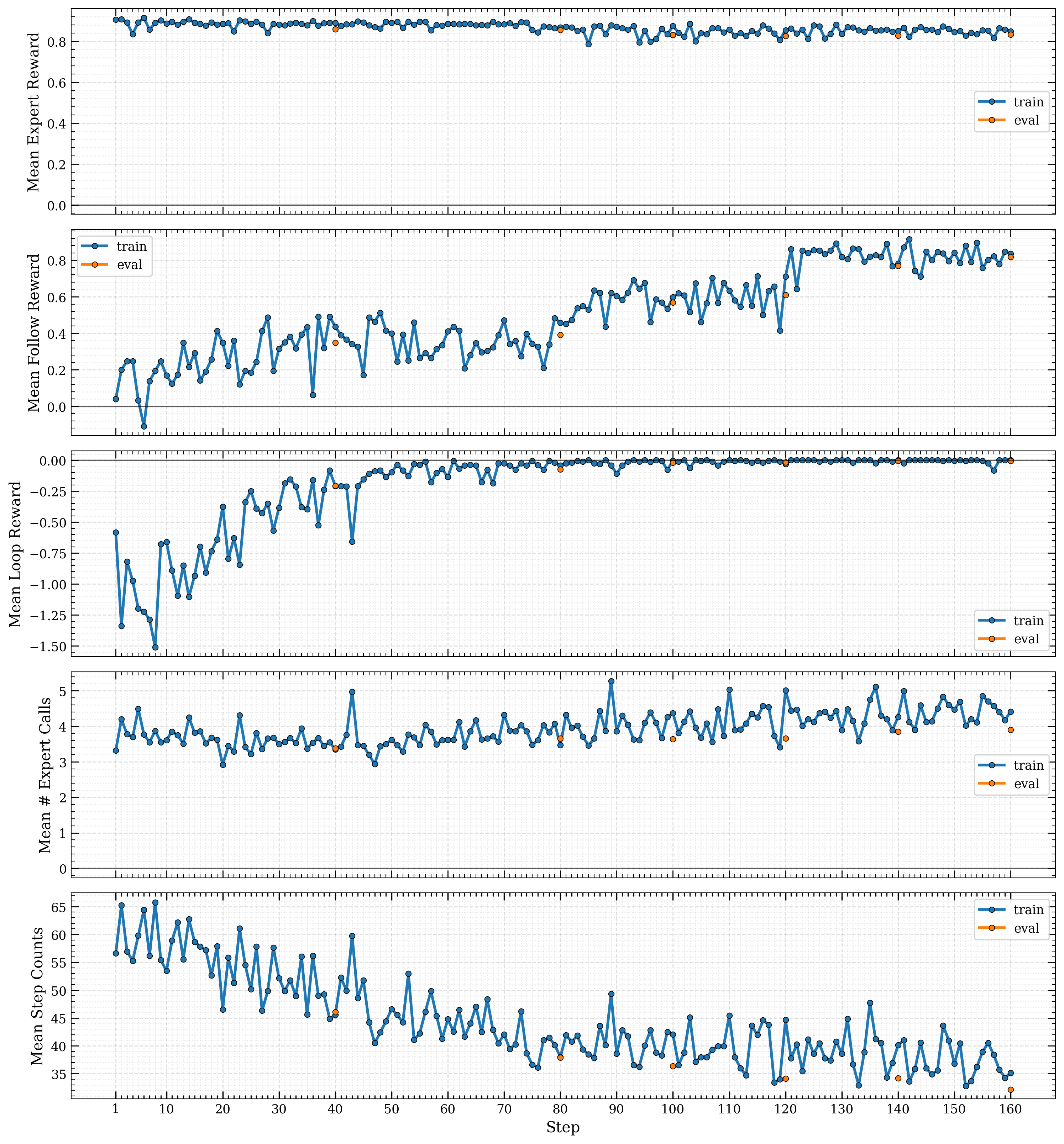}
\caption{Phase~II training logs starting from the SFT checkpoint, which include mean rewards for selected components.}
\label{fig:mean-rl-reward}
\end{figure}

We include additional diagnostics from Phase~II training in Fig.~\ref{fig:mean-rl-reward}.
Mean \textit{Expert} reward remains highly stable throughout training, staying within a narrow band of approximately \(0.82\)–\(0.88\) across all steps, indicating that expert responses consistently satisfy the reward criterion and are not degraded by later-stage shaping.
In contrast, \textit{Follow} reward exhibits a clear upward trend.
During the first stage (steps 1--80), it fluctuates substantially, ranging roughly from \(0.2\) to \(0.5\).
After step~80, following the transition to follow-aggressive shaping, it increases steadily and stabilizes in the \(0.75\)–\(0.85\) range by step~160.
\textit{Loop} reward shows the most dramatic change: it begins near \(-1.5\), rises rapidly during the loop-first shaping phase, and approaches \(0.0\) by step~80.
Thereafter it remains close to zero with minimal variance, indicating that degenerative looping behavior is largely eliminated and does not re-emerge in later training.
The mean number of expert calls stays relatively stable at approximately \(3.5\)–\(4.5\) calls per episode, with a mild upward drift in later stages, suggesting that performance gains are not driven by increased expert reliance.
Finally, mean step counts decrease substantially over training, dropping from roughly \(55\)–\(65\) steps early on to about \(35\)–\(40\) steps by the end of training, consistent with more direct trajectories and reduced redundant actions.

\begin{figure}[t]
\centering
\includegraphics[width=\linewidth]{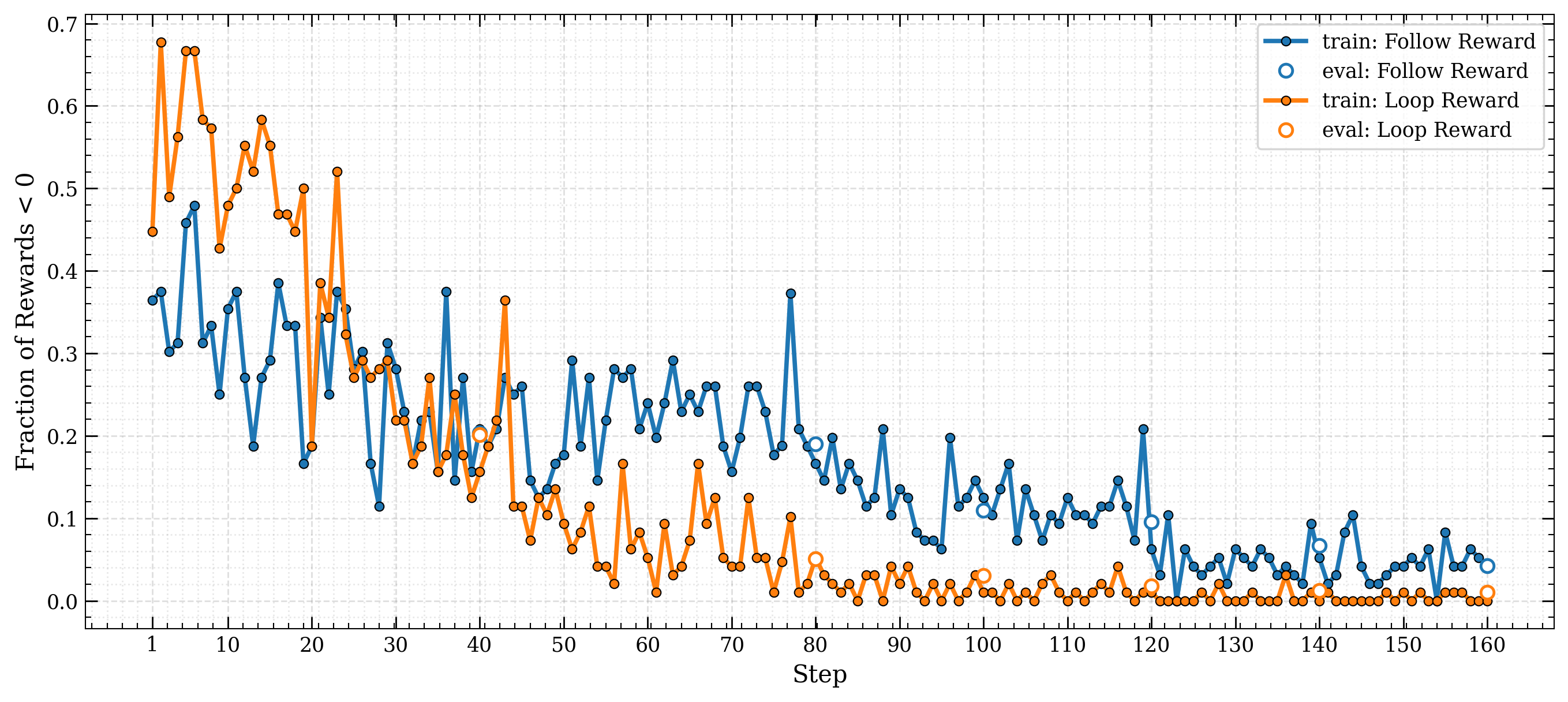}
\caption{Fraction of trajectories with \textit{Loop} and \textit{Follow} reward $\leq$0 during Phase~II training.}
\label{fig:negative-rl-reward}
\end{figure}

Fig.~\ref{fig:negative-rl-reward} further characterizes training dynamics by reporting the fraction of trajectories whose \textit{Follow} or \textit{Loop} reward is non-positive.
Early in training, a substantial fraction of trajectories incur negative rewards for both components.
Within the first 20--30 steps, approximately \(40\%\)–\(65\%\) of trajectories receive non-positive \textit{Loop} reward, while roughly \(25\%\)–\(40\%\) exhibit non-positive \textit{Follow} reward.
As training progresses, both fractions decline steadily.
The fraction of trajectories with non-positive \textit{Loop} reward drops sharply during the loop-first shaping phase and falls below \(10\%\) by around step~50, approaching zero shortly thereafter and remaining near zero for the remainder of training.
The fraction of trajectories with non-positive \textit{Follow} reward decreases more gradually, falling below \(10\%\) by approximately step~100 and stabilizing in the \(3\%\)–\(8\%\) range by the end of training.
Train and evaluation curves closely track each other throughout, indicating that these behavioral improvements generalize beyond the training trajectories.

Fig.~\ref{fig:loops} analyzes degenerative looping behavior during Phase~I, by measuring the fraction of trajectories that contain repetitive tool-use sequences of length greater than \(L\).
Our evaluation here also includes an active intervention mechanism that is activated at \(L=15\) for the initial looping sequence detected, and \(L=8\) for subsequent detections.
Across all agents, the fraction decreases monotonically as \(L\) increases, but with substantial differences in tail behavior.
Baseline agents exhibit long repetitive sequences with high frequency: the SWE-agent-LM-32B baseline retains approximately \(15\%\)–\(20\%\) of trajectories with repetition length exceeding \(L=10\), and remains above \(10\%\) even at \(L=40\).
In contrast, all \sys-7B variants substantially suppress long repetitive sequences.
For \sys-7B, the fraction of trajectories with repetition length exceeding \(L=10\) drops below \(10\%\), and falls to near zero by \(L \approx 15\), indicating that extended tool-use loops are largely captured and eliminated properly by the active intervention mechanism; however, as we showed in Fig.~\ref{fig:minicoder_variants}, this is still ineffective in improving SWE-bench performance. 
Among \sys variants, differences across expert backends (not shown in the figure for clarity) are minor and primarily affect short repetition lengths, while the long-tail behavior remains consistently suppressed.
To construct these curves, for each trajectory we compute the maximum length of any consecutive identical tool-use action run, and then for each threshold \(L \in [0,40]\) we plot the fraction of trajectories whose maximum run length exceeds \(L\) (i.e., \(\mathbb{E}[\mathbb{1}[\max\_{{\rm run}} > L]]\) over trajectories).
Shaded regions indicate \(95\%\) bootstrap uncertainty bands obtained by resampling trajectories within each group (400 resamples; fixed random seed for determinism) and evaluating the same indicator-based curve on each resample.

\section{Expert-augmented Synthetic Trajectory Generation}
\label{app:synthetic-data}

We use Claude Sonnet~3.7 to generate full expert-augmented trajectories following the procedure described in \S\ref{subsec:sft}, using the same underlying task dataset as SWE-smith~\citep{swe_smith_dataset,yang2025swe}.
Tasks are randomly sampled, and trajectories are generated with rejection sampling until we obtain 4.8K accepted trajectories, approximately matching the training set size used in SWE-smith.
Because the exact task composition may differ due to random sampling, we perform a controlled check by regenerating trajectories \emph{without} expert augmentation and applying SFT to the same baseline model (Qwen-2.5-Coder-Instruct, 7B and 32B variants) using an identical 4.8K-sample budget.
The resulting SWE-bench performance closely matches that reported for SWE-smith, indicating that differences observed in later experiments are not attributable to dataset size or sampling effects.
The prompt used to generate expert-augmented trajectories is shown below in Fig.~\ref{fig:dataset-gen-prompt}. 
Importantly, the expert used in synthetic data generation is Claude Sonnet 3.7 itself; the system prompt fed to it is described in Fig.~\ref{fig:expert-prompt-synthetic-swe-smith-gen}. The exact expert mechanism is described in \S\ref{app:expert-details}. Note that, apart from synthetic data generation, the gold patch and unit tests are never provided to the expert.

\begin{minipage}{\linewidth}

\begin{example}
\small
You are a helpful assistant that can interact with a computer to solve tasks. \\

Call \texttt{ask\_expert\_llm} to help guide your thinking, before/after localizing the issue, applying a patch, when considering edge cases, and whenever you have trouble reproducing the error (though not always necessary). \\

\noindent\textbf{Instance template.} \\
\texttt{<uploaded\_files>} \\
\texttt{\{\{working\_dir\}\}} \\
\texttt{</uploaded\_files>} \\
I've uploaded a python code repository in the directory \texttt{\{\{working\_dir\}\}}. Consider the following PR description: \\

\texttt{<pr\_description>} \\
\texttt{\{\{problem\_statement\}\}} \\
\texttt{</pr\_description>} \\

\texttt{<context>} \\
\texttt{\{\{context\}\}} \\
\texttt{</context>} \\

Can you help me implement the necessary changes to the repository so that the requirements specified in the \texttt{<pr\_description>} are met? \\
I've already taken care of all changes to any of the test files described in the \texttt{<pr\_description>}. This means you \textbf{DON'T} have to modify the testing logic or any of the tests in any way! \\
Your task is to make the minimal changes to non-tests files in the \texttt{\{\{working\_dir\}\}} directory to ensure the \texttt{<pr\_description>} is satisfied. \\

Follow these steps to resolve the issue: \\
1. As a first step, it might be a good idea to find and read code relevant to \texttt{<pr\_description>}. Call the \texttt{ask\_expert\_llm} tool before or after doing this as needed. \\
2. Create a script to reproduce the error and execute it with \texttt{python <filename.py>} using the bash tool, to confirm the error. Call the \texttt{ask\_expert\_llm} tool before or after doing this as needed. \\
3. Edit the source code of the repo to resolve the issue. Call the \texttt{ask\_expert\_llm} tool before or after doing this as needed. \\
4. Rerun your reproduce script and confirm that the error is fixed! \\
5. Think about edge cases and make sure your fix handles them as well. Call the \texttt{ask\_expert\_llm} tool before or after doing this as needed. \\

Your thinking should be thorough and so it's fine if it's very long. Use \texttt{ask\_expert\_llm} in moderation---avoid consecutive calls, and always integrate its response into your reasoning (e.g., ``Based on the expert's advice''). \\
\end{example}
\captionof{figure}{
System prompt used for expert-augmented trajectory generation.
This is a slight modification of the standard SWE-agent prompt, adding explicit guidance for when to invoke \texttt{ask\_expert\_llm} and how to incorporate expert feedback, while keeping the underlying task structure unchanged.
}
\label{fig:dataset-gen-prompt}

\end{minipage}

\begin{minipage}{\linewidth}
\begin{example}
\small
You are a helpful assistant that can interact with a computer to solve tasks. \\

\texttt{<IMPORTANT>} \\
\(\bullet\) If the user provides a path, you should \textbf{NOT} assume it's relative to the current working directory. Instead, explore the file system to find the file before working on it. \\
\(\bullet\) Call \texttt{ask\_expert\_llm} to help guide your thinking, e.g., before/after localizing the issue, before/after applying a patch, when thinking through edge cases, and whenever you encounter problems reproducing the error. \\
When you call \texttt{ask\_expert\_llm}, treat it as a multi-turn collaborator: ask for specific, actionable steps; then execute those steps; then report back what you did and what you observed before asking again. \\
\texttt{</IMPORTANT>} \\[0.5em]

You have access to the following functions: \\[0.25em]
\texttt{---- BEGIN FUNCTION \#1: bash ----} \\
\texttt{Description: Execute a bash command in the terminal.} \\
\texttt{Parameters:} \\
\hspace*{1em}\texttt{(1) command (string, required): The bash command to execute. Can be empty to view additional logs when previous exit code is -1. Can be ctrl+c to interrupt the currently running process.} \\
\texttt{---- END FUNCTION \#1 ----} \\[0.25em]

\texttt{<....omitted for clarity...>} \\[0.5em]

\texttt{---- BEGIN FUNCTION \#4: ask\_expert\_llm ----} \\
\texttt{Description: Call ask\_expert\_llm to help guide your thinking, e.g., when thinking through edge cases.} \\
\texttt{Parameters:} \\
\hspace*{1em}\texttt{(1) question (string, required): The expert question, e.g., Do you have any advice on how to proceed?} \\
\hspace*{1em}\texttt{(2) budget\_tokens (integer, optional): Token budget hint if applicable} \\
\texttt{---- END FUNCTION \#4 ----} \\[0.5em]

\texttt{<....omitted for clarity...>} \\[0.5em]

Can you help me implement the necessary changes to the repository so that the requirements specified in the \texttt{<pr\_description>} are met? \\
I've already taken care of all changes to any of the test files described in the \texttt{<pr\_description>}. This means you \textbf{DON'T} have to modify the testing logic or any of the tests in any way! \\
Your task is to make the minimal changes to non-tests files in the \texttt{\{\{working\_dir\}\}} directory to ensure the \texttt{<pr\_description>} is satisfied. \\

Follow these steps to resolve the issue: \\
1. As a first step, it might be a good idea to find and read code relevant to \texttt{<pr\_description>}. Call the \texttt{ask\_expert\_llm} tool before or after doing this as needed. \\
2. Create a script to reproduce the error and execute it with \texttt{python <filename.py>} using the bash tool, to confirm the error. Call the \texttt{ask\_expert\_llm} tool before or after doing this as needed. \\
3. Edit the source code of the repo to resolve the issue. Call the \texttt{ask\_expert\_llm} tool before or after doing this as needed. \\
4. Rerun your reproduce script and confirm that the error is fixed! \\
5. Think about edge cases and make sure your fix handles them as well. Call the \texttt{ask\_expert\_llm} tool before or after doing this as needed. \\

Your thinking should be thorough and so it's fine if it's very long. Use \texttt{ask\_expert\_llm} in moderation---avoid consecutive calls, and always integrate its response into your reasoning (e.g., ``Based on the expert's advice''). \\
\end{example}
\captionof{figure}{
Prompt used by \sys agents.
This prompt is a slight modification of the standard SWE-agent prompt: it (i) makes \texttt{ask\_expert\_llm} usage explicit and structured as a multi-turn collaborator, (ii) adds a path-resolution instruction to avoid incorrect assumptions about file locations, and (iii) exposes the tool/function-call interface used during rollouts, while preserving the underlying task and repository setup.
}
\label{fig:sys-agent-prompt}
\end{minipage}

\section{Expert Tool Details}
\label{app:expert-details}

The expert prompt used by Claude Sonnet 3.7 when generating expert-augmented trajectories is shown in Fig.~\ref{fig:expert-prompt-synthetic-swe-smith-gen}. The standard expert prompt (used everywhere else) is shown in Fig.~\ref{fig:expert-prompt-standard}. There are two prompts used by our expert judge: the first is used to judge whether the call to the expert was appropriate (Fig.~\ref{fig:expert-judge-prompt}), and the second is an expert judge call to obtain the \textit{Follow} score (Fig.~\ref{fig:follow-judge-prompt}). Finally, the prompt used by \sys SLM agents is shown in Fig.~\ref{fig:sys-agent-prompt}. 

\noindent\textbf{Expert tool implementation (SWE-agent instantiation).}
Our ``expert'' is implemented using standard agent tooling: an external advice function (\texttt{ask\_expert\_llm}) exposed to the agent alongside regular tools (e.g., \texttt{bash} and \texttt{submit}). Concretely, SWE-agent loads this tool via a tool-bundle entry (the \texttt{tools/expert\_llm} bundle) in the agent YAML configuration.

\noindent\textbf{Context passed to the expert.}
When invoked, \texttt{ask\_expert\_llm} reads the problem statement and a recent window of the agent's conversation from SWE-agent’s exported history files inside the container. The tool constructs a single structured context block containing (i) the task description and (ii) a JSON dump of recent agent messages (after history processing), and appends the agent’s explicit question. This context is provided as background (not as the expert’s own prior turns), and the expert response is returned as an observation wrapped in \texttt{<expert\_llm\_guidance>} tags.

\noindent\textbf{Expert routing, limits, and optional scoring.}
The expert model can be configured independently of the main agent (e.g., via \texttt{EXPERT\_MODEL} and \texttt{EXPERT\_OPENAI\_API\_BASE}), enabling a separate endpoint/model for expert calls. To control cost, the tool enforces a per-task call quota (default 6). 

\noindent\textbf{Preventing expert advice from being elided by history processing.}
In SWE-agent, for SWE-agent-LM 7B/32B models, the model input context is derived from the raw interaction history via a configurable chain of \texttt{history\_processors}. A commonly used default processor (\texttt{last\_n\_observations}, often with \(n{=}5\)) elides older tool observations by replacing them with a short ``(lines omitted)'' placeholder to control context length. In our SWE-smith instantiation, we slightly modify this behavior so that observations containing the \texttt{<expert\_llm\_guidance>} marker are never elided by \texttt{last\_n\_observations}. This ensures the agent continues to see the expert’s advice beyond the last-\(n\) window, preventing it from disappearing after a handful of turns.

\noindent\textbf{RL-oriented expert supervision: warranted-score, follow-through, and a post-hoc terminal judge.}
To support reinforcement learning with expert-augmented trajectories, we extend the standard \texttt{ask\_expert\_llm} tool call with two auxiliary judging mechanisms and log their outputs into the trajectory metadata for reward shaping. First, when \texttt{ASK\_EXPERT\_LLM\_ENABLE\_SCORING=1}, the expert is prompted to return strict JSON containing both textual guidance and a continuous \texttt{warranted\_score}\(\in[0,1]\) indicating whether escalation was justified; the tool returns only the guidance to the agent but records the score for training-time penalties/rewards. Second, when \texttt{ASK\_EXPERT\_LLM\_ENABLE\_FOLLOW\_JUDGE=1}, \texttt{ask\_expert\_llm} issues an additional hidden judge call that scores adherence to the \emph{previous} expert guidance (call \(k{-}1\)) based on the subsequent agent messages up to the current expert invocation (call \(k\)), producing \texttt{follow\_score}\(\in[0,1]\) that is logged as \texttt{expert\_follow\_score} (not shown to the agent). A key semantic consequence is that the \emph{final} expert guidance call has no later call to “carry” its follow-through score; therefore, in our modified SkyRL training harness we add an out-of-loop, post-hoc follow judge that replays the same follow-judge logic after rollout completion and writes a separate \texttt{final\_expert\_follow\_score} field into \texttt{trajectory.info} (used by the RL reward without altering per-call expert penalties). These expert-derived scores are then weighted in the composite reward (e.g., via \texttt{generator.expert\_cfg.weight} and \texttt{generator.follow\_cfg.weight} in the SkyRL launcher).

\begin{minipage}{\linewidth}
\begin{example}
\small
You are an expert software engineer solving SWE-bench--style tasks. \\

Provide the most direct and precise guidance to the agent’s question. \\
Treat any provided conversation or repository details strictly as background context---they are \emph{not} your prior messages. \\

You are also given a ground-truth patch. \\
This patch represents the correct solution: do \textbf{not} copy it verbatim, but use it to direct the agent toward the correct solution as clearly and quickly as possible. \\

If the agent’s framing is inaccurate, restate the real issue clearly and redirect them to the correct solution. \\
Keep your answer concise, actionable, and technically accurate. \\
\end{example}
\captionof{figure}{
System prompt used for Claude Sonnet~3.7 acting as the expert during expert-augmented trajectory generation.
The expert is instructed to provide high-level corrective guidance grounded in a hidden ground-truth patch, without revealing the solution verbatim.
}
\label{fig:expert-prompt-synthetic-swe-smith-gen}
\end{minipage}

\begin{minipage}{\linewidth}
\begin{example}
\small
You are an expert software engineer assisting an agent solving SWE-bench--style tasks. \\

Provide clear, direct, and technically precise guidance that helps the agent make concrete progress (e.g., reproducing the bug, isolating the faulty logic, or implementing a minimal fix). \\
You may have strong intuitions about the root cause or solution, but focus on communicating actionable steps the agent should take in the next few steps. \\

Treat any provided conversation or repository details strictly as background context---they are the agent's own prior messages, not yours. \\
If you believe the agent is on the wrong track (e.g., it has not found the right file, diagnosis, fix direction, or even the right question), redirect it to the correct line of investigation.
More generally, respond as you would if you were solving the task at this moment. \\

Keep your answer concise, actionable, and technically accurate. \\
As the agent is solving SWE-bench--style tasks, do not mention pull requests, commit messages, or other GitHub workflow artifacts. \\
\end{example}
\captionof{figure}{
System prompt used by the expert model under our standard expert-collaboration setting.
The expert is instructed to provide concise, actionable guidance and to correct the agent when it diverges from the most promising debugging path.
}
\label{fig:expert-prompt-standard}
\end{minipage}

\begin{minipage}{\linewidth}
\begin{example}
\small
\noindent\textit{(Expert-judge extension; appended to the expert prompt)} \\

Return \textbf{ONLY} a single valid JSON object (no Markdown, no extra text) with exactly the following schema: \\

\texttt{\{} \\
\hspace*{1em}\texttt{"guidance": string,} \\
\hspace*{1em}\texttt{"warranted\_score": float,} \\
\texttt{\}} \\

\noindent where \texttt{warranted\_score} is in \([0.0, 1.0]\): \\
\hspace*{1em}\(\bullet\) Values below \(0.5\) indicate that the expert call was generally unnecessary. \\
\hspace*{1em}\(\bullet\) A value of \(1.0\) indicates the expert call was clearly necessary and well-timed. \\
\hspace*{1em}\(\bullet\) Be strict in your scoring. \\[0.25em]

\noindent\textbf{Example:} \\
\texttt{\{} \\
\hspace*{1em}\texttt{"guidance": "Do X...\textbackslash nThen verify Y...",} \\
\hspace*{1em}\texttt{"warranted\_score": 0.8} \\
\texttt{\}} \\[0.25em]

\noindent\textbf{Important:} \\
If \texttt{guidance} spans multiple lines, encode newlines as \texttt{\textbackslash n}. \\
\end{example}
\captionof{figure}{
Expert-judge prompt used to evaluate whether an expert call was warranted.
This prompt is appended to the standard expert prompt (Fig.~\ref{fig:expert-prompt-standard}) and instructs the judge to both provide corrective guidance and assign a strict scalar score indicating whether the agent’s expert query was appropriate.
}
\label{fig:expert-judge-prompt}
\end{minipage}

\begin{minipage}{\linewidth}
\begin{example}
\small
\noindent\textbf{System prompt (follow judge).} \\

You are evaluating whether an agent followed prior expert guidance. \\
Return \textbf{ONLY} a single valid JSON object (no Markdown, no extra text) with exactly the following schema: \\

\texttt{\{} \\
\hspace*{1em}\texttt{"follow\_score": float,} \\
\hspace*{1em}\texttt{"rationale": string} \\
\texttt{\}} \\

\noindent where \texttt{follow\_score} is in \([0.0, 1.0]\): \\
\hspace*{1em}\(\bullet\) Values below \(0.5\) indicate that the agent generally did not follow prior expert guidance. \\
\hspace*{1em}\(\bullet\) Be strict in your scoring. \\[0.25em]

\noindent\textbf{Scoring guidance.} \\
If the agent was instructed to perform a specific action (e.g., view a file and apply an edit) but only partially followed it (e.g., viewed without editing), this should receive a low score.
Requesting clarification or assistance in order to complete a specific instructed action should be considered as following that guidance for that action.
If there is no clear instruction to follow, set \texttt{follow\_score} to \texttt{null}. \\[0.5em]

\noindent\textbf{User input to the judge.} \\

\texttt{<prior\_expert\_guidance>} \\
\texttt{<expert guidance text>} \\
\texttt{</prior\_expert\_guidance>} \\

\texttt{<agent\_followup\_conversation>} \\
\texttt{<subsequent agent messages (JSON-encoded)>} \\
\texttt{</agent\_followup\_conversation>} \\

\texttt{<agent\_current\_question\_to\_expert>} \\
\texttt{<current agent question>} \\
\texttt{</agent\_current\_question\_to\_expert>} \\
\end{example}
\captionof{figure}{
Follow-judge prompt used to assess whether the agent followed prior expert guidance.
The judge receives the previous expert advice, the agent’s subsequent interaction trace, and the agent’s current expert query, and outputs a strict scalar score indicating adherence to guidance along with a brief rationale.
}
\label{fig:follow-judge-prompt}
\end{minipage}

\section{Phase II Reward Modeling and Shaping Details}
\label{app:phase2-reward}

This appendix defines the reward terms referenced in Phase~II (\S\ref{subsec:rl}). The goal is to make (i) degenerative stalling and
(ii) non-collaborative expert usage explicitly suboptimal, while keeping rewards stable for RL post-training.

\subsection{Correctness and Similarity Terms}
\label{app:phase2-correct-sim}

These apply to the code patch (output artifact) generated by the coding agent. Both terms are computed similarly to~\citep{copet2025cwm}.

\noindent\textbf{Correctness.}
Let \(\tau\) denote the full agent interaction trajectory (interleaved sequence of agent thoughts/actions and tool responses), and let \(p(\tau)\) denote the final patch the agent submits at the end of \(\tau\).
We compute \(R_{\text{correct}}(\tau,x)\in\{0,1\}\) by re-evaluating \(p(\tau)\) in a
fresh environment identical to the task backend used for rollout generation:
\[
R_{\text{correct}}(\tau,x)=\mathbb{1}\{p(\tau)\ \text{passes verification}\}.
\]
A patch is considered correct (resolved) if all unit tests pass (e.g., as in SWE-Bench).
Otherwise, it is considered unresolved.
Here, \(x\) denotes the task instance metadata (e.g., the dataset record), which includes the gold patch used below and any information needed to re-run verification.

\noindent\textbf{Similarity fallback (unresolved only).}
When unresolved (\(R_{\text{correct}}=0\)), we compute a similarity score
\(\mathrm{sim}(\tau,x)\in[0,1]\) between the model patch and a gold patch after filtering diff noise.
Concretely, we drop diffs for newly-added files (e.g., ``\texttt{new file mode}'' blocks or diffs involving \texttt{/dev/null}) and drop the git noise line \texttt{\textbackslash No newline at end of file}, then compute string similarity on the remaining unified-diff text.
With threshold \(\theta\) (we use \(\theta=0.5\) in practice),
\[
R_{\text{sim}}(\tau,x)=
\begin{cases}
0, & \mathrm{sim}(\tau,x)\ge \theta,\\
-1, & \mathrm{sim}(\tau,x)< \theta.
\end{cases}
\]
This discrete fallback discourages unrelated patches while preserving reward stability. We use a thresholded signal to avoid instability under policy updates and prevent similarity from dominating optimization.
Intuitively, when a trajectory fails verification, we still want a stable learning signal: clearly unrelated patches are penalized (\(-1\)), while patches that substantially overlap the gold fix receive a neutral score (\(0\)).
This can improve learning on difficult instances where the policy cannot yet produce a fully test-passing patch, but can still learn to localize and edit the right region.
If the model patch is missing (or becomes empty after filtering), we treat similarity as \(0\) and assign \(R_{\text{sim}}=-1\). 

\subsection{Stall Penalty}
\label{app:phase2-loop}

\noindent\textbf{Base-command extraction.}
To detect stalled progress (often manifesting as action loops), we convert each tool/action step into a \emph{base command} and form a sequence \(c_1,\dots,c_M\).
We normalize commands to make superficial variants count as the same action, then measure repetition on the normalized sequence.
Our normalization mirrors the reward implementation:
\begin{itemize}
\item Strip leading environment-variable assignments (e.g., \texttt{FOO=bar python -m pytest} \(\mapsto\) \texttt{python -m pytest}).
\item For chained shell commands joined by \texttt{\&\&} or \texttt{;}, keep the \emph{last} subcommand (e.g., \texttt{cd repo \&\& pytest} \(\mapsto\) \texttt{pytest}).
\item Define the base command as the first token (e.g., \texttt{pytest}, \texttt{ls}), except for a small set treated as two-word bases (e.g., \texttt{git status}, \texttt{git diff}, \texttt{str\_replace\_editor view}).
\item Collapse ``navigation-like'' operations into one equivalence class so mixed navigation streaks count as repetition: e.g., \texttt{grep}, \texttt{find}, and \texttt{str\_replace\_editor view} are treated as identical for stall detection.
\end{itemize}
For example, repeated \texttt{git diff} actions are counted as identical, and a sequence like \texttt{grep} \(\to\) \texttt{find} \(\to\) \texttt{str\_replace\_editor view} is treated as a repeated navigation streak.

Let \(s_1,s_2,\dots\) denote lengths of maximal consecutive identical-command streaks \emph{in temporal order} as we scan the trajectory (i.e., \(s_1\) is the first streak length, \(s_2\) the next streak after the command changes, etc.).

\noindent\textbf{Triggered, capped stall penalty.}
We use a triggered rule: the penalty only activates once any streak is long enough; after the first trigger,
subsequent streaks use a lower threshold \(k_2<k_1\). The charged exceedance matches the implementation, which uses a small ``\(+1\)'' offset so that a streak that just reaches the threshold is already penalized:
\begin{align}
R_{\text{loop}}(\tau)
=
\max\Big(
c_{\text{loop}},\;
-\lambda_{\text{loop}}
\Big(
\max(0,s_1-k_1+1)+\sum_{j\ge 2}\max(0,s_j-k_2+1)
\Big)
\Big),
\label{eq:rloop}
\end{align}
where \(\lambda_{\text{loop}}>0\) and \(c_{\text{loop}}<0\) caps the magnitude.
Operationally, the first term contributes when \(s_1\ge k_1\); after the first trigger, later streaks contribute when \(s_j\ge k_2\) for \(j\ge 2\).
This makes stalling \emph{sparse but decisive}: short repeats are tolerated, while true degeneracy is sharply penalized.

\subsection{Expert-as-Judge Collaboration Terms}
\label{app:phase2-expert-judge}

These terms train the agent to collaborate with the expert in a multi-turn, pair-programming-like manner: they discourage unnecessary escalation, and they directly supervise the agent’s behavior \emph{after} receiving advice (executing the requested steps and reporting back).

Let \(\mathcal{E}(\tau)=\{e_i\}_{i=1}^{N}\) denote the ordered set of expert-call events (as in \S\ref{subsec:rl}), where each event
\(e_i=(t_i,q_i,g_i,\tilde{s}_i)\) records the timestep \(t_i\), the agent query \(q_i\) sent to the expert, the expert guidance \(g_i\), and the compact context packet \(\tilde{s}_i\) provided to the expert.
In our implementation, \(\tilde{s}_i\) is a compact \emph{bundle of recent context}, not a single message: it includes the problem statement and a truncated tail of recent processed conversation messages (a fixed-size window for token control) from \(\tau\), serialized for the expert.
Concretely, we pass (at most) the last \(\approx 10\) processed messages, then drop the leading system prompt and the first user message that repeats the full problem statement, and remove action/thought fields from these messages before serialization.
For follow-through, we associate each guidance \(g_i\) with the subsequent response segment \(\Delta_i(\tau)\), i.e., the agent’s behavior after receiving \(g_i\) up to and including the next expert call (or trajectory end if there is no subsequent call).

We log two judge scores computed by the prompts shown earlier in this appendix (warrant judge: Fig.~\ref{fig:expert-judge-prompt}; follow judge: Fig.~\ref{fig:follow-judge-prompt}):
\[
u_i := J_{\text{warrant}}(q_i,\tilde{s}_i)\in[0,1],
\qquad
f_i := J_{\text{follow}}(g_i,\Delta_i(\tau))\in[0,1].
\]
In the logged trajectories, these scores appear as per-call numeric fields (warrant: \texttt{expert\_score}; follow: \texttt{expert\_follow\_score}) that the reward function consumes directly.
Note that online follow judging stores the follow score on the \emph{next} expert call (evaluating adherence to the previous guidance), so the final guidance may optionally contribute an additional terminal follow score computed post-hoc.

\noindent\textbf{Invocation quality (\(R_{\text{expert}}\)).}
We convert the per-call warrant scores \(\{u_i\}\) into a single trajectory-level term that discourages low-value escalation while keeping scale stable as the number of expert calls \(N\) varies:
\[
\phi(u)=
\begin{cases}
u, & u \ge \tau_{\text{low}},\\
p_{\text{low}}, & u < \tau_{\text{low}},
\end{cases}
\qquad p_{\text{low}}\le 0,
\qquad
R_{\text{warrant}}(\tau)=\mathrm{Agg}\big(\{\phi(u_i)\}_{i=1}^{N}\big),
\]
where \(\mathrm{Agg}\) is typically \(\text{mean}\) (or \(\text{min}\) for stricter budgeting).
Here, \(u\) is a dummy variable used to define the scalar transform \(\phi(\cdot)\), which is then applied to each \(u_i\).

To prevent expert spam, let \(n_{\text{b2b}}\) be the number of back-to-back expert calls. We add a capped penalty:
\[
R_{\text{b2b}}(\tau)=\max(-1,\;\lambda_{\text{b2b}}\,n_{\text{b2b}}),
\qquad \lambda_{\text{b2b}}\le 0.
\]
Optionally, with an expert budget \(Q\), let \(n_{\text{over}}=\max(0,N-Q)\) and
\(R_{\text{quota}}(\tau)=p_{\text{over}}\,n_{\text{over}}\) with \(p_{\text{over}}\le 0\).
We do \emph{not} enable this quota term in our runs; it is included for completeness.
We combine:
\[
R_{\text{expert}}(\tau)=R_{\text{warrant}}(\tau)+R_{\text{b2b}}(\tau)+\mathbb{1}[\text{quota enabled}]\,R_{\text{quota}}(\tau).
\]

\noindent\textbf{Follow-through (\(R_{\text{follow}}\)).}
We shape follow-through similarly:
\[
\psi(f)=
\begin{cases}
f, & f \ge \tau_{\text{follow}},\\
p_{\text{follow-low}}, & f < \tau_{\text{follow}},
\end{cases}
\qquad p_{\text{follow-low}}\le 0,
\qquad
R_{\text{follow}}(\tau)=\mathrm{Agg}\big(\{\psi(f_i)\}_{i\in\mathcal{I}}\big),
\]
where \(\mathcal{I}\) indexes events for which follow-through is defined (e.g., excluding terminal calls).

\subsection{Gating Functions}
\label{app:phase2-gates}

We gate only the auxiliary term \(R_{\text{other}}\) to prevent correctness or similarity
from compensating for pathological behavior; loop and follow-through terms are always applied:
\[
g_{\text{loop}}(\tau)=
\begin{cases}
0, & R_{\text{loop}}(\tau)\le a_2,\\
0.5, & R_{\text{loop}}(\tau)\le a_1,\\
1, & \text{otherwise},
\end{cases}
\qquad
g_{\text{follow}}(\tau)=
\begin{cases}
0, & R_{\text{follow}}(\tau)\le b_2,\\
0.5, & R_{\text{follow}}(\tau)\le b_1,\\
1, & \text{otherwise},
\end{cases}
\]
for thresholds \(a_2<a_1\le 0\) and \(b_2<b_1\le 0\). The main text uses
\(R_{\text{total}} = R_{\text{loop}} + w_{\text{follow}}R_{\text{follow}} + g_{\text{loop}}g_{\text{follow}}R_{\text{other}}\)
(Eq.~\ref{eq:rtotal-main}). Notably, \(R_{\text{follow}}\) is never gated out; gating applies solely
to \(R_{\text{other}}\), while loop and follow-through penalties remain active throughout training.
In our reward-shaping schedule below, we apply the follow gate only in Stage~II by setting \(g_{\text{follow}}\equiv 1\) during Stage~I (i.e., the follow gate is effectively inactive in Stage~I).

\subsection{Reward Shaping Schedule}
\label{app:phase2-shaping}

While Phase~I SFT teaches the SLM to imitate expert interaction patterns, it does not reliably induce the behaviors required for effective collaboration.
In particular, we observe two persistent failure modes: (i) the model fails to escalate when progress stalls, leading to long degenerative loops; and (ii) even when advice is obtained, the model often fails to follow it or to report back appropriately.
We therefore apply a two-stage shaping curriculum that progressively tightens constraints to first suppress stalling and then enforce follow-through.

\noindent\textbf{Stage I: Loop aggressive shaping (escalation induction).}
We make stalling strongly suboptimal by (i) increasing loop/stall penalty magnitude via a more negative cap \(c_{\text{loop}}\) and/or larger
\(\lambda_{\text{loop}}\) in Eq.~\ref{eq:rloop}, and (ii) setting the loop gate \(g_{\text{loop}}\) to downweight \(R_{\text{other}}\)
more aggressively. In this stage, we keep follow-through shaping mild:
\(w_{\text{follow}}\) is small and we keep the follow gate inactive (i.e., \(g_{\text{follow}}\equiv 1\)).
This reliably shifts the policy from ``repeat failed actions'' to ``seek help when stuck.''

\noindent\textbf{Stage II: Loop+follow aggressive shaping (pair programming).}
After stalling is largely suppressed, we increase the strength of follow-through shaping by
(i) increasing \(w_{\text{follow}}\),
(ii) setting \(p_{\text{follow-low}}\) more negative, and/or
(iii) making \(g_{\text{follow}}\) more aggressive (lower thresholds \(b_1,b_2\)),
so that failing to execute and report back on expert guidance becomes strongly suboptimal.
In addition, Stage~II enforces a hard no-expert gate: trajectories with zero expert calls
incur a fixed negative penalty, making unguided execution explicitly suboptimal.
This stage converts one-shot escalation into multi-turn collaboration.

\noindent\textbf{Reporting.}
In \S\ref{sec:results}, we report ablations over this schedule (Stage~I only vs.\ Stage~I+II) and show that Stage~I primarily
reduces stagnation while Stage~II improves expert-guidance adherence and end-to-end solve rate.
All concrete hyperparameters (weights, caps, thresholds) are provided in \S\ref{sec:exp-setup}.

\clearpage

\section{Example Trajectory Snippets}
\label{app:trajectory-snippets}

For completeness, we include partial, representative trajectory snippets (left to right) corresponding to the three examples in Fig.~\ref{fig:arch}(b) for the SWE-bench Verified instance \texttt{sympy\_\_sympy-14976}.
In the first example (SWE-agent-LM-7B; Fig.~\ref{fig:traj-snippet-sweagent-7b-lm}), the SLM stalls by repeatedly reading different segments of \texttt{lambdify.py}, forming tentative hypotheses, and continuing localization until it aborts due to runtime limits (we omit the remaining steps for brevity).
In the second example (\sys-7B P1, Sonnet 3.7; Fig.~\ref{fig:traj-snippet-protege-p1}), the SLM requests expert help but fails to follow the expert’s instruction to focus on the \texttt{lambdify} function in \texttt{/testbed/sympy/utilities/lambdify.py} (though it manages to follow the suggestion to inspect \texttt{MPMATH\_TRANSLATIONS}). It then stalls by repeatedly issuing similar commands to view different portions of the \texttt{Rational} class, rather than re-engaging the expert.
Finally, in the third example (\sys-7B P2, Sonnet 3.7; Fig.~\ref{fig:traj-snippet-protege-p2}), the SLM follows the expert’s instructions end-to-end, reports progress back, and ultimately converges to the correct fault localization in \texttt{pycode.py} via this collaboration.

\definecolor{trajthoughtcolor}{HTML}{1f77b4}
\definecolor{trajactioncolor}{HTML}{ff7f0e}
\newtcolorbox{trajthoughtbox}{
  before skip=0.1em,
  after skip=0.1em,
  colback=trajthoughtcolor!5!white,
  colframe=trajthoughtcolor,
  boxrule=0.4pt,
  arc=1pt,
  left=1mm,
  right=1mm,
  top=0.8mm,
  bottom=0.8mm
}
\newtcolorbox{trajactionbox}{
  before skip=0.1em,
  after skip=0.1em,
  colback=trajactioncolor!4!white,
  colframe=trajactioncolor!80!black,
  boxrule=0.4pt,
  arc=1pt,
  left=1mm,
  right=1mm,
  top=0.8mm,
  bottom=0.8mm
}

\noindent
\textbf{(a) SWE-agent-7B-LM.} Instance \texttt{sympy\_\_sympy-14976}.

\begin{trajthoughtbox}
\small
Let's continue reading the function:\\
{\ttfamily str\_replace\_editor view /testbed/sympy/utilities/lambdify.py  --view\_range 250 300}
\end{trajthoughtbox}
\begin{trajactionbox}
\small\ttfamily
Here's the result of running `cat -n` on /testbed/sympy/utilities/lambdify.py:\\
   250	        values or numpy arrays as arguments.  However, in some cases\\
   251	        the generated function relies on the input being a numpy array:\\
   252	\\
   253	        >>> from sympy import Piecewise\\
\textit{[Additional output omitted for brevity.]}
\end{trajactionbox}
\begin{trajthoughtbox}
\small
Let's continue reading the function:\\
{\ttfamily str\_replace\_editor view /testbed/sympy/utilities/lambdify.py  --view\_range 300 350}
\end{trajthoughtbox}
\begin{trajactionbox}
\small\ttfamily
Here's the result of running `cat -n` on /testbed/sympy/utilities/lambdify.py:\\
   300	    Matrix([[1, 3]])\\
   301	\\
   302	    Tuple arguments are handled and the lambdified function should\\
   303	    be called with the same type of arguments as were used to create\\
\textit{[Additional output omitted for brevity.]}
\end{trajactionbox}
\begin{trajthoughtbox}
\small
Let's continue reading the function:\\
{\ttfamily str\_replace\_editor view /testbed/sympy/utilities/lambdify.py  --view\_range 350 400}
\end{trajthoughtbox}
\begin{trajactionbox}
\small\ttfamily
Here's the result of running `cat -n` on /testbed/sympy/utilities/lambdify.py:\\
   350	    array([[ 1.,  2.],\\
   351	           [ 3.,  4.]], dtype=float32)\\
   352	\\
   353	    """\\
\textit{[Additional output omitted for brevity.]}
\end{trajactionbox}
\begin{trajthoughtbox}
\small
Let's continue reading the function:\\
{\ttfamily str\_replace\_editor view /testbed/sympy/utilities/lambdify.py  --view\_range 400 450}
\end{trajthoughtbox}
\begin{trajactionbox}
\small\ttfamily
Here's the result of running `cat -n` on /testbed/sympy/utilities/lambdify.py:\\
   400	            from sympy.printing.pycode import MpmathPrinter as Printer\\
   401	        elif \_module\_present('numpy', namespaces):\\
   402	            from sympy.printing.pycode import NumPyPrinter as Printer\\
   403	        elif \_module\_present('numexpr', namespaces):\\
\textit{[Additional output omitted for brevity.]}
\end{trajactionbox}
\begin{trajthoughtbox}
\small
Let's continue reading the function:\\
{\ttfamily str\_replace\_editor view /testbed/sympy/utilities/lambdify.py  --view\_range 450 500}
\end{trajthoughtbox}
\begin{trajactionbox}
\small\ttfamily
Here's the result of running `cat -n` on /testbed/sympy/utilities/lambdify.py:\\
   450	\\
   451	    funcname = '\_lambdifygenerated'\\
   452	\\
   453	    if \_module\_present('tensorflow', namespaces):\\
\textit{[Additional output omitted for brevity.]}
\end{trajactionbox}
\begin{trajthoughtbox}
\small
Now let's look at the `\_TensorflowEvaluatorPrinter` and `\_EvaluatorPrinter` classes which are responsible for generating the code for different modules:\\
{\ttfamily str\_replace\_editor view /testbed/sympy/printing/lambdarepr.py}
\end{trajthoughtbox}
\begin{trajactionbox}
\small\ttfamily
Here's the result of running `cat -n` on /testbed/sympy/printing/lambdarepr.py:\\
     1	from \_\_future\_\_ import print\_function, division\\
     2	from distutils.version import LooseVersion as V\\
     3	\\
     4	from .str import StrPrinter\\
\textit{[Additional output omitted for brevity.]}
\end{trajactionbox}
\begin{trajthoughtbox}
\small
Let's also look at the PycodePrinter class which is likely responsible for generating the code for the 'mpmath' module:\\
{\ttfamily str\_replace\_editor view /testbed/sympy/printing/pycode.py  --view\_range 1 50}
\end{trajthoughtbox}
\begin{trajactionbox}
\small\ttfamily
Here's the result of running `cat -n` on /testbed/sympy/printing/pycode.py:\\
     1	"""\\
     2	Python code printers\\
     3	\\
     4	This module contains python code printers for plain python as well as NumPy \& SciPy enabled code.\\
\textit{[Additional output omitted for brevity.]}
\end{trajactionbox}
\begin{trajthoughtbox}
\small
Let's look at the MpmathPrinter class which is specifically for the 'mpmath' module:\\
{\ttfamily grep -r "MpmathPrinter" /testbed}
\end{trajthoughtbox}
\begin{trajactionbox}
\small\ttfamily
/testbed/sympy/printing/pycode.py:class MpmathPrinter(PythonCodePrinter):\\
/testbed/sympy/printing/pycode.py:for k in MpmathPrinter.\_kf:\\
/testbed/sympy/printing/pycode.py:    setattr(MpmathPrinter, '\_print\_\%s' \% k, \_print\_known\_func)\\
/testbed/sympy/printing/pycode.py:    setattr(MpmathPrinter, '\_print\_\%s' \% k, \_print\_known\_const)\\
/testbed/sympy/printing/tests/test\_pycode.py:    MpmathPrinter, NumPyPrinter, PythonCodePrinter, pycode, SciPyPrinter\\
\textit{[Additional output omitted for brevity.]}
\end{trajactionbox}
\begin{trajthoughtbox}
\small
Let's examine the MpmathPrinter class:\\
{\ttfamily grep -A 50 "class MpmathPrinter" /testbed/sympy/printing/pycode.py}
\end{trajthoughtbox}
\begin{trajactionbox}
\small\ttfamily
class MpmathPrinter(PythonCodePrinter):\\
    """\\
    Lambda printer for mpmath which maintains precision for floats\\
    """\\
    printmethod = "\_mpmathcode"\\
\textit{[Additional output omitted for brevity.]}
\end{trajactionbox}
\begin{trajthoughtbox}
\small
Let's also look at the `\_known\_functions\_mpmath` dictionary:\\
{\ttfamily grep -A 20 "\_known\_functions\_mpmath" /testbed/sympy/printing/pycode.py}
\end{trajthoughtbox}
\begin{trajactionbox}
\small\ttfamily
\_known\_functions\_mpmath = dict(\_in\_mpmath, **\{\\
    'sign': 'sign',\\
\})\\
\_known\_constants\_mpmath = \{\\
    'Pi': 'pi'\\
\textit{[Additional output omitted for brevity.]}
\end{trajactionbox}
\begin{trajthoughtbox}
\small
Let's see where `\_known\_functions\_mpmath` is defined:\\
{\ttfamily grep -r "\_known\_functions\_mpmath =" /testbed}
\end{trajthoughtbox}
\begin{trajactionbox}
\small\ttfamily
/testbed/sympy/printing/pycode.py:\_known\_functions\_mpmath = dict(\_in\_mpmath, **\{
\end{trajactionbox}
\begin{trajthoughtbox}
\small
Let's see what `\_in\_mpmath` contains:\\
{\ttfamily grep -A 20 "\_in\_mpmath" /testbed/sympy/printing/pycode.py}
\end{trajthoughtbox}
\begin{trajactionbox}
\small\ttfamily
\_not\_in\_mpmath = 'log1p log2'.split()\\
\_in\_mpmath = [(k, v) for k, v in \_known\_functions\_math.items() if k not in \_not\_in\_mpmath]\\
\_known\_functions\_mpmath = dict(\_in\_mpmath, **\{\\
    'sign': 'sign',\\
\})\\
\textit{[Additional output omitted for brevity.]}
\end{trajactionbox}
\begin{trajthoughtbox}
\small
Let's see where `\_known\_functions\_math` is defined:\\
{\ttfamily grep -r "\_known\_functions\_math =" /testbed}
\end{trajthoughtbox}
\begin{trajactionbox}
\small\ttfamily
/testbed/sympy/printing/pycode.py:\_known\_functions\_math = \{
\end{trajactionbox}
\begin{trajthoughtbox}
\small
Let's see the definition of `\_known\_functions\_math`:\\
{\ttfamily str\_replace\_editor view /testbed/sympy/printing/pycode.py  --view\_range 1 50}
\end{trajthoughtbox}
\begin{trajactionbox}
\small\ttfamily
Here's the result of running `cat -n` on /testbed/sympy/printing/pycode.py:\\
     1	"""\\
     2	Python code printers\\
     3	\\
     4	This module contains python code printers for plain python as well as NumPy \& SciPy enabled code.\\
\textit{[Additional output omitted for brevity.]}
\end{trajactionbox}
\begin{trajthoughtbox}
\small
Let's continue reading the definition of `\_known\_functions\_math`:\\
{\ttfamily str\_replace\_editor view /testbed/sympy/printing/pycode.py  --view\_range 50 100}
\end{trajthoughtbox}
\begin{trajactionbox}
\small\ttfamily
Here's the result of running `cat -n` on /testbed/sympy/printing/pycode.py:\\
    50	    'log2': 'log2',\\
    51	    'sin': 'sin',\\
    52	    'sinh': 'sinh',\\
    53	    'Sqrt': 'sqrt',\\
\textit{[Additional output omitted for brevity.]}
\end{trajactionbox}
\begin{trajthoughtbox}
\small
Let's continue reading the PythonCodePrinter class:\\
{\ttfamily str\_replace\_editor view /testbed/sympy/printing/pycode.py  --view\_range 100 150}
\end{trajthoughtbox}
\begin{trajactionbox}
\small\ttfamily
Here's the result of running `cat -n` on /testbed/sympy/printing/pycode.py:\\
   100	        super(PythonCodePrinter, self).\_\_init\_\_(settings)\\
   101	        self.module\_imports = defaultdict(set)\\
   102	        self.known\_functions = dict(self.\_kf, **(settings or \{\}).get(\\
   103	            'user\_functions', \{\}))\\
\textit{[Additional output omitted for brevity.]}
\end{trajactionbox}
\vspace{6pt}
\captionof{figure}{
Example trajectory snippet from SWE-agent-7B-LM, instance \texttt{sympy\_\_sympy-14976}.
We omit repeated runtime warnings and truncate long tool outputs for readability.
}
\label{fig:traj-snippet-sweagent-7b-lm}

\vspace{1.0em}
\noindent
\textbf{(b) SWE-Prot\'eg\'e-7B P1 (Sonnet 3.7).} Instance \texttt{sympy\_\_sympy-14976}.

\begin{trajthoughtbox}
\small
{\ttfamily
ask\_expert\_llm 'Based on the PR description, there seems to be an issue with mpmath not properly handling rational numbers in lambdified expressions. Looking at the code in /testbed/sympy/utilities/lambdify.py, where should I focus my attention to fix this issue?'}
\end{trajthoughtbox}
\begin{trajactionbox}
\small\ttfamily
<expert\_llm\_guidance>\\
The issue is in how rational numbers are handled when lambdifying expressions with the mpmath module. Based on the PR description, when a rational number like `S(1)/3` appears in an expression that's lambdified with `modules='mpmath'`, it remains as a Python fraction (`232/3`) rather than being converted to an mpmath object with full precision.\\
\\
You should focus on these specific areas:\\
\\
1. The `lambdify` function in `/testbed/sympy/utilities/lambdify.py` (around line 166)\\
2. The `MPMATH\_TRANSLATIONS` dictionary (around line 45)\\
\\
The key issue is that when rational numbers appear in expressions lambdified with mpmath, they need to be converted to mpmath's high-precision representation rather than remaining as Python fractions.\\
\\
The solution likely involves modifying how rational numbers are handled in the code generation process for mpmath specifically. You'll need to ensure that when the mpmath module is used, rational numbers like `S(1)/3` are converted to `mpmath.mpf(1)/mpmath.mpf(3)` or an equivalent mpmath representation that preserves full precision.\\
\\
This would fix the issue where `nsolve` gets reduced precision results because the rational constant isn't evaluated at full precision.\\
</expert\_llm\_guidance>
\end{trajactionbox}
\begin{trajthoughtbox}
\small
Based on the expert's guidance, I need to focus on how rational numbers are handled when lambdifying expressions with the mpmath module. Let's examine the relevant parts of the code:\\
{\ttfamily str\_replace\_editor view /testbed/sympy/utilities/lambdify.py  --view\_range 45 71}
\end{trajthoughtbox}
\begin{trajactionbox}
\small\ttfamily
Here's the result of running `cat -n` on /testbed/sympy/utilities/lambdify.py:\\
    45	MPMATH\_TRANSLATIONS = \{\\
    46	    "Abs": "fabs",\\
    47	    "elliptic\_k": "ellipk",\\
    48	    "elliptic\_f": "ellipf",\\
\textit{[Additional output omitted for brevity.]}
\end{trajactionbox}
\begin{trajthoughtbox}
\small
Now let's look at how rational numbers are handled in general:\\
{\ttfamily grep -r "S\textbackslash{}(" /testbed/sympy/core/}
\end{trajthoughtbox}
\begin{trajactionbox}
\small\ttfamily
grep: Unmatched ( or \textbackslash{}(
\end{trajactionbox}
\begin{trajthoughtbox}
\small
Let's try a different approach:\\
{\ttfamily grep -r "S(" /testbed/sympy/core/}
\end{trajthoughtbox}
\begin{trajactionbox}
\small\ttfamily
/testbed/sympy/core/numbers.py:    >>> mod\_inverse(S(2)/7, S(5)/2)\\
/testbed/sympy/core/numbers.py:                    return (S.NegativeOne**num[0]*num[1]*S(2)**num[2])
\\
/testbed/sympy/core/numbers.py:    >>> S('.[3]')  \# repeating digits in brackets\\
/testbed/sympy/core/numbers.py:    >>> S('3**2/10')  \# general expressions\\
/testbed/sympy/core/numbers.py:        >>> (S(-3)/2).as\_content\_primitive()\\
\textit{[Additional output omitted for brevity.]}
\end{trajactionbox}
\begin{trajthoughtbox}
\small
Let's look for files related to rational numbers:\\
{\ttfamily find /testbed -type f -name "*.py" | grep -i "number" | head -10}
\end{trajthoughtbox}
\begin{trajactionbox}
\small\ttfamily
/testbed/sympy/polys/numberfields.py\\
/testbed/sympy/polys/tests/test\_numberfields.py\\
/testbed/sympy/functions/combinatorial/numbers.py\\
/testbed/sympy/functions/combinatorial/tests/test\_comb\_numbers.py\\
/testbed/sympy/core/numbers.py\\
\textit{[Additional output omitted for brevity.]}
\end{trajactionbox}
\begin{trajthoughtbox}
\small
Let's examine the `numbers.py` file which likely contains the rational number implementation:\\
{\ttfamily str\_replace\_editor view /testbed/sympy/core/numbers.py  --view\_range 1 50}
\end{trajthoughtbox}
\begin{trajactionbox}
\small\ttfamily
Here's the result of running `cat -n` on /testbed/sympy/core/numbers.py:\\
     1	from \_\_future\_\_ import print\_function, division\\
     2	\\
     3	import decimal\\
     4	import fractions\\
\textit{[Additional output omitted for brevity.]}
\end{trajactionbox}
\begin{trajthoughtbox}
\small
Let's look at the rational number implementation:\\
{\ttfamily grep -n "class Rational" /testbed/sympy/core/numbers.py}
\end{trajthoughtbox}
\begin{trajactionbox}
\small\ttfamily
1388:class Rational(Number):\\
2469:class RationalConstant(Rational):
\end{trajactionbox}
\begin{trajthoughtbox}
\small
Let's look at the Rational class:\\
{\ttfamily str\_replace\_editor view /testbed/sympy/core/numbers.py  --view\_range 1388 1450}
\end{trajthoughtbox}
\begin{trajactionbox}
\small\ttfamily
Here's the result of running `cat -n` on /testbed/sympy/core/numbers.py:\\
  1388	class Rational(Number):\\
  1389	    """Represents rational numbers (p/q) of any size.\\
  1390	\\
  1391	    Examples\\
\textit{[Additional output omitted for brevity.]}
\end{trajactionbox}
\begin{trajthoughtbox}
\small
Let's continue looking at the Rational class:\\
{\ttfamily str\_replace\_editor view /testbed/sympy/core/numbers.py  --view\_range 1450 1500}
\end{trajthoughtbox}
\begin{trajactionbox}
\small\ttfamily
Here's the result of running `cat -n` on /testbed/sympy/core/numbers.py:\\
  1450	\\
  1451	    Access numerator and denominator as .p and .q:\\
  1452	\\
  1453	    >>> r = Rational(3, 4)\\
\textit{[Additional output omitted for brevity.]}
\end{trajactionbox}
\begin{trajthoughtbox}
\small
Let's continue looking at the Rational class:\\
{\ttfamily str\_replace\_editor view /testbed/sympy/core/numbers.py  --view\_range 1500 1550}
\end{trajthoughtbox}
\begin{trajactionbox}
\small\ttfamily
Here's the result of running `cat -n` on /testbed/sympy/core/numbers.py:\\
  1500	                    p = p.replace(' ', '')\\
  1501	                    pq = p.rsplit('/', 1)\\
  1502	                    if len(pq) == 2:\\
  1503	                        p, q = pq\\
\textit{[Additional output omitted for brevity.]}
\end{trajactionbox}
\begin{trajthoughtbox}
\small
Let's continue looking at the Rational class:\\
{\ttfamily str\_replace\_editor view /testbed/sympy/core/numbers.py  --view\_range 1550 1600}
\end{trajthoughtbox}
\begin{trajactionbox}
\small\ttfamily
Here's the result of running `cat -n` on /testbed/sympy/core/numbers.py:\\
  1550	        obj = Expr.\_\_new\_\_(cls)\\
  1551	        obj.p = p\\
  1552	        obj.q = q\\
  1553	        return obj\\
\textit{[Additional output omitted for brevity.]}
\end{trajactionbox}

\captionof{figure}{
Example trajectory snippet from \sys-7B P1 (Sonnet 3.7), instance \texttt{sympy\_\_sympy-14976}.
We show the agent thought and the corresponding tool output for each thought/action pair; long tool outputs are truncated, while expert guidance blocks are shown in full.
}
\label{fig:traj-snippet-protege-p1}

\vspace{1.0em}
\noindent
\textbf{(c) \sys-7B P2 (Sonnet 3.7).} Instance \texttt{sympy\_\_sympy-14976\_0}.

\begin{trajthoughtbox}
\small
Let's look at the specific part of the code that handles the 'mpmath' module, since that's where the issue is occurring according to the PR description:\\
{\ttfamily
ask\_expert\_llm 'In the Sympy lambdify.py file, where would the handling of rational numbers specifically for the '"'"'mpmath'"'"' module be implemented? I'"'"'m looking for the code that processes expressions when using modules='"'"'mpmath'"'"' and why rationals aren'"'"'t being properly wrapped.'}
\end{trajthoughtbox}
\begin{trajactionbox}
\small\ttfamily
<expert\_llm\_guidance>\\
You need to look at how rational numbers are handled in the lambdify function. The issue is that when using the 'mpmath' module, rational numbers like 232/3 aren't being converted to mpmath's high-precision format.\\
\\
First, examine the lambdify function implementation in detail, located in:\\
/testbed/sympy/utilities/lambdify.py \\
\\
Then look at how the printer handles different types of expressions, also located in:\\
/testbed/sympy/utilities/lambdify.py \\
\\
The issue is likely in how rational numbers are printed when the mpmath module is selected. Look for code that handles Rational objects or fractions, and check if there's special handling for the mpmath module that's missing. You'll probably need to modify the \_EvaluatorPrinter class to ensure that rational numbers are converted to mpmath.mpf() calls when the mpmath module is used.\\
</expert\_llm\_guidance>
\end{trajactionbox}
\begin{trajthoughtbox}
\small
Let me try again with the correct format:\\
{\ttfamily str\_replace\_editor view /testbed/sympy/utilities/lambdify.py  --view\_range 704 744}
\end{trajthoughtbox}
\begin{trajactionbox}
\small\ttfamily
Here's the result of running `cat -n` on /testbed/sympy/utilities/lambdify.py:\\
   704	    def \_preprocess(self, args, expr):\\
   705	        """Preprocess args, expr to replace arguments that do not map\\
   706	        to valid Python identifiers.\\
   707	\\
\textit{[Additional output omitted for brevity.]}
\end{trajactionbox}
\begin{trajthoughtbox}
\small
Now let's look at how subexpressions are processed, which is likely where the rational numbers should be handled:\\
{\ttfamily str\_replace\_editor view /testbed/sympy/utilities/lambdify.py  --view\_range 746 763}
\end{trajthoughtbox}
\begin{trajactionbox}
\small\ttfamily
Here's the result of running `cat -n` on /testbed/sympy/utilities/lambdify.py:\\
   746	    def \_subexpr(self, expr, dummies\_dict):\\
   747	        from sympy.matrices import DeferredVector\\
   748	        from sympy import sympify\\
   749	\\
\textit{[Additional output omitted for brevity.]}
\end{trajactionbox}
\begin{trajthoughtbox}
\small
Let's also check how functions are printed, particularly focusing on the mpmath module:\\
{\ttfamily str\_replace\_editor view /testbed/sympy/utilities/lambdify.py  --view\_range 654 688}
\end{trajthoughtbox}
\begin{trajactionbox}
\small\ttfamily
Here's the result of running `cat -n` on /testbed/sympy/utilities/lambdify.py:\\
   654	    def doprint(self, funcname, args, expr):\\
   655	        """Returns the function definition code as a string."""\\
   656	        from sympy import Dummy\\
   657	\\
\textit{[Additional output omitted for brevity.]}
\end{trajactionbox}
\begin{trajthoughtbox}
\small
Now, let's create a script to reproduce the issue mentioned in the PR description:\\
{\ttfamily str\_replace\_editor create /testbed/reproduce\_issue.py --file\_text '\#!/usr/bin/env python3\\
from sympy import symbols, Eq, rf, S, Float, nsolve\\
import inspect\\
\# Define the variable\\
x = symbols('"'"'x'"'"')\\
\# Create the equation from the PR description\\
eqn = Eq(rf(18, x), 77 + S(1)/3)\\
\# Print the equation\\
print("Original equation:", eqn)\\
\# Lambdify the expression with mpmath module\\
f = lambda x: eval(inspect.getsource(lambda y: (rf(18, y) - (77 + S(1)/3)).evalf())[10:-2])\\
print("\\textbackslash{}nLambda function source:")\\
print(inspect.getsource(f))\\
\# Test with nsolve\\
x0 = nsolve(eqn, Float('"'"'1.5'"'"', 64), prec=64)\\
print("\\textbackslash{}nSolution x0:", x0)\\
print("rf(18, x0).evalf(64):", rf(18, x0).evalf(64))\\
print("Expected: \\textasciitilde{}77.3333333333333")'}
\end{trajthoughtbox}
\begin{trajactionbox}
\small\ttfamily
File created successfully at: /testbed/reproduce\_issue.py
\end{trajactionbox}
\begin{trajthoughtbox}
\small
Let's run the script to see the issue:\\
{\ttfamily cd /testbed \&\& python reproduce\_issue.py}
\end{trajthoughtbox}
\begin{trajactionbox}
\small\ttfamily
/testbed/sympy/solvers/diophantine.py:3188: SyntaxWarning: "is" with a literal. Did you mean "=="?\\
  if feasible is 1:  \# it's prime and k == 2\\
Original equation: Eq(RisingFactorial(18, x), 232/3)\\
Lambda function source:\\
f = lambda x: eval(inspect.getsource(lambda y: (rf(18, y) - (77 + S(1)/3)).evalf())[10:-2])\\
\textit{[Additional output omitted for brevity.]}
\end{trajactionbox}
\begin{trajthoughtbox}
\small
Now I understand the issue. When using `lambdify(modules='mpmath')`, the rational number `232/3` is not being properly wrapped in mpmath's high-precision format. Let's ask the expert for guidance on how to fix this:\\
{\ttfamily
ask\_expert\_llm 'I'"'"'ve identified that the issue is in the \_EvaluatorPrinter class in lambdify.py. When using modules='"'"'mpmath'"'"', rational numbers like 232/3 aren'"'"'t being converted to mpmath.mpf() calls. How should I modify the code to ensure that rational numbers are properly wrapped in mpmath.mpf() when the mpmath module is selected?'}
\end{trajthoughtbox}
\begin{trajactionbox}
\small\ttfamily
<expert\_llm\_guidance>\\
You're on the right track. The issue is that when using the mpmath module, rational numbers need to be wrapped in mpmath.mpf() calls to maintain precision.\\
\\
Let's look at the \_EvaluatorPrinter class which handles how expressions are printed in the lambdify function:\\
\\
1. First, check the \_EvaluatorPrinter class implementation, located in:\\
/testbed/sympy/utilities/lambdify.py\\
\\
2. Then look for how rational numbers are handled in the printer. There should be a method like \_print\_Rational or similar.\\
\\
3. The fix will likely involve modifying how Rational objects are printed when the module is 'mpmath'. You'll need to wrap the rational in mpmath.mpf() instead of just using Python's division operator.\\
\\
After identifying the specific method, modify it to check if the module is 'mpmath' and if so, wrap the rational in mpmath.mpf().\\
</expert\_llm\_guidance>
\end{trajactionbox}

\begin{trajthoughtbox}
\small
Let's examine the \_EvaluatorPrinter class more closely to find how rational numbers are handled:\\
{\ttfamily str\_replace\_editor view /testbed/sympy/utilities/lambdify.py  --view\_range 400 450}
\end{trajthoughtbox}
\begin{trajactionbox}
\small\ttfamily
Here's the result of running `cat -n` on /testbed/sympy/utilities/lambdify.py:\\
   400	            from sympy.printing.pycode import MpmathPrinter as Printer\\
   401	        elif \_module\_present('numpy', namespaces):\\
   402	            from sympy.printing.pycode import NumPyPrinter as Printer\\
   403	        elif \_module\_present('numexpr', namespaces):\\
\textit{[Additional output omitted for brevity.]}
\end{trajactionbox}
\begin{trajthoughtbox}
\small
Now let's look for how rational numbers are handled in the printers. Based on the PR description, we need to focus on the `\_EvaluatorPrinter` class which seems to handle the actual printing of expressions for evaluation:\\
{\ttfamily grep -r "\_print\_Rational" /testbed/sympy/}
\end{trajthoughtbox}
\begin{trajactionbox}
\small\ttfamily
/testbed/sympy/polys/numberfields.py:    def \_print\_Rational(self, expr):\\
/testbed/sympy/polys/numberfields.py:        return "mpi('\\
grep: /testbed/sympy/polys/\_\_pycache\_\_/numberfields.cpython-39.pyc: binary file matches\\
/testbed/sympy/printing/ccode.py:    def \_print\_Rational(self, expr):\\
/testbed/sympy/printing/mathml.py:    def \_print\_Rational(self, e):\\
\textit{[Additional output omitted for brevity.]}
\end{trajactionbox}
\begin{trajthoughtbox}
\small
Let's check if there's a specific implementation of `\_EvaluatorPrinter` in the lambdify module:\\
{\ttfamily grep -A 10 "\_EvaluatorPrinter" /testbed/sympy/utilities/lambdify.py}
\end{trajthoughtbox}
\begin{trajactionbox}
\small\ttfamily
        funcprinter = \_EvaluatorPrinter(printer, dummify)\\
    funcstr = funcprinter.doprint(funcname, args, expr)\\
    funclocals = \{\}\\
    global \_lambdify\_generated\_counter\\
    filename = '<lambdifygenerated-\%s>' \% \_lambdify\_generated\_counter\\
\textit{[Additional output omitted for brevity.]}
\end{trajactionbox}
\begin{trajthoughtbox}
\small
Now I'll look at the `\_print\_Rational` method in the `\_EvaluatorPrinter` class. Since we don't see it explicitly defined there, it likely inherits it from the parent class. Let's check the base class:\\
{\ttfamily
ask\_expert\_llm 'The \_EvaluatorPrinter class in lambdify.py doesn'"'"'t seem to have a \_print\_Rational method defined. It inherits from LambdaPrinter. Where should I look for the implementation of how rational numbers are handled during lambdification, especially for the mpmath module case?'}
\end{trajthoughtbox}
\begin{trajactionbox}
\small\ttfamily
<expert\_llm\_guidance>\\
You need to look at the MpmathPrinter class since that's what's used when modules='mpmath'. Let's examine the implementation:\\
\\
1. First, check the MpmathPrinter in sympy/printing/pycode.py:\\
\\
2. Then look for the \_print\_Rational method in this class or its parent classes. If MpmathPrinter doesn't override \_print\_Rational, check its parent class (likely PythonCodePrinter):\\
\\
3. The issue is likely that the \_print\_Rational method in PythonCodePrinter (or whatever parent class) is not properly wrapping rational numbers in mpmath.mpf() when used with the mpmath module. You'll need to modify the MpmathPrinter class to override \_print\_Rational to use mpmath.mpf() for proper precision.\\
</expert\_llm\_guidance>
\end{trajactionbox}
\begin{trajthoughtbox}
\small
Based on the expert's advice, I need to check the MpmathPrinter class in the pycode.py file:\\
{\ttfamily str\_replace\_editor view /testbed/sympy/printing/pycode.py  --view\_range 1 100}
\end{trajthoughtbox}
\begin{trajactionbox}
\small\ttfamily
Here's the result of running `cat -n` on /testbed/sympy/printing/pycode.py:\\
     1	"""\\
     2	Python code printers\\
     3	\\
     4	This module contains python code printers for plain python as well as NumPy \& SciPy enabled code.\\
\textit{[Additional output omitted for brevity.]}
\end{trajactionbox}
\begin{trajthoughtbox}
\small
Let's search for the MpmathPrinter class and the \_print\_Rational method:\\
{\ttfamily grep -n "class MpmathPrinter" /testbed/sympy/printing/pycode.py}
\end{trajthoughtbox}
\begin{trajactionbox}
\small\ttfamily
314:class MpmathPrinter(PythonCodePrinter):
\end{trajactionbox}

\begin{trajthoughtbox}
\small
(no thought logged)\\
{\ttfamily grep -n "\_print\_Rational" -A 5 /testbed/sympy/printing/pycode.py | grep -A 5 MpmathPrinter}
\end{trajthoughtbox}
\begin{trajactionbox}
\small\ttfamily
Your command ran successfully and did not produce any output.
\end{trajactionbox}

\begin{trajthoughtbox}
\small
Let's look at the MpmathPrinter class implementation:\\
{\ttfamily str\_replace\_editor view /testbed/sympy/printing/pycode.py  --view\_range 314 350}
\end{trajthoughtbox}
\begin{trajactionbox}
\small\ttfamily
Here's the result of running `cat -n` on /testbed/sympy/printing/pycode.py:\\
   314	class MpmathPrinter(PythonCodePrinter):\\
   315	    """\\
   316	    Lambda printer for mpmath which maintains precision for floats\\
   317	    """\\
\textit{[Additional output omitted for brevity.]}
\end{trajactionbox}

\captionof{figure}{
Example trajectory snippet from SWE-Prot\'eg\'e-7B P2 (Sonnet 3.7), instance \texttt{sympy\_\_sympy-14976\_0}.
We omit repeated runtime warnings, truncate long tool outputs, and show expert guidance blocks in full.
}
\label{fig:traj-snippet-protege-p2} 

\end{document}